\documentclass[11pt,a4paper]{article}
\pdfoutput=1
\usepackage{jheppub}
\usepackage{gensymb}
\usepackage{subfigure}
\usepackage{amssymb,amsmath}
\usepackage{graphicx}
\usepackage{color}
\usepackage{cancel}
\usepackage[colorlinks=true
,urlcolor=blue
,citecolor=blue
,linkcolor=blue
,pagecolor=blue
,linktocpage=true
,pdfproducer=medialab
]{hyperref}
\usepackage[section]{placeins}
 \usepackage[numbers]{natbib}
\usepackage{notoccite}
\makeatletter \renewcommand{\@dotsep}{10000} \makeatother
\def\be{\begin{equation}}
\def\ee{\end{equation}}
\def\bea{\begin{eqnarray}}
\def\eea{\end{eqnarray}}
\def\bi{\begin{itemize}}
\def\ei{\end{itemize}}

\usepackage[nodisplayskipstretch]{setspace}

\def\mgut{M_{\rm GUT}}

\newcommand{\beq}{\begin{equation}}
\newcommand{\eeq}{\end{equation}}

\newcommand*{\Scale}[2][4]{\scalebox{#1}{$#2$}}%

\newcommand{\mo}{\textsc{MicrOMEGAs}}
\begin{document}

\begin{titlepage}
\pagestyle{empty}

\vspace*{0.2in}
\begin{center}
{\Large \bf    Model Characterization and Dark Matter in the Secluded $U(1)^\prime$ Model}\\
\vspace{1cm}
{\bf  Ya\c{s}ar Hi\c{c}y\i lmaz$^{a,b}$\footnote{Email: yasarhicyilmaz@balikesir.edu.tr },
Levent Selbuz$^{c,}$\footnote{Email: selbuz@eng.ankara.edu.tr},
Levent Solmaz$^{a,}$\footnote{Email: lsolmaz@balikesir.edu.tr }  {\rm and}
Cem Salih $\ddot{\rm U}$n$^{d,e,}\hspace{0.05cm}$\footnote{E-mail: cemsalihun@uludag.edu.tr}}
\vspace{0.5cm}

{\it
$^a$Department of Physics, Bal\i kesir University, TR10145, Bal\i kesir, Turkey, \\
$^b$School of Physics $\&$ Astronomy, University of Southampton, Highfield, Southampton SO17 1BJ,UK, \\
$^c$Department of Engineering Physics, Ankara
University, TR06100 Ankara, Turkey, \\
$^d$Department of Physics, Bursa Uluda\~{g} University, TR16059 Bursa, Turkey, \\
$^e$Departamento de Ciencias Integradas y Centro de Estudios Avanzados en F\'{i}sica Matem\'aticas y Computación, Campus del Carmen, Universidad de Huelva, Huelva 21071, Spain 
}
\end{center}

$^\bigstar$ This paper is dedicated to the memory of Prof. Levent Solmaz, who unexpectedly passed away on August 16, 2021.

\vspace{0.1cm}
\begin{abstract}
	
\noindent {\small We consider a class of $U(1)^\prime$-extended MSSM in which the $U(1)^{\prime}$ symmetry is broken by vacuum expectation values (VEVs) of four MSSM singlet fields. While one MSSM singlet field interacts with the MSSM Higgs fields, three of them interact only with each other in forming a secluded sector. Assigning universal $U(1)^{\prime}$ charges for three families, the anomaly cancellation condition requires exotic fields which are assumed to be heavy and decoupled. We discuss a variety of $U(1)^{\prime}$ charge assignments and anomaly cancellation, $Z^{\prime}/Z$ hierarchy, neutralinos, charginos and the Higgs sector. We find that the typical spectra involve two CP-odd Higgs bosons lighter than about 200 GeV and 600 GeV respectively, which are mostly formed by the MSSM singlet fields. If the relic density of dark matter is saturated only by a neutralino, compatible solutions predict LSP neutralinos formed by the MSSM singlet fields in the mass scales below about 600 GeV, while it is possible to realize MSSM neutralino LSP above these mass scales. One can classify the implications in three scenarios. Scenario I involves NLSP charginos, while Scenario II involves charginos which do not participate coannihilation processes with the LSP neutralino. These two scenarios predict MSSM singlet LSP neutralinos, while Scenario I leads to larger scattering cross-sections of dark matter. Scenario III has the solutions in which the MSSM neutralinos are considerably involved in LSP decomposition which yields very large scattering cross-section excluded by the direct detection experiments.}

\end{abstract}

\end{titlepage}

\tableofcontents

\noindent \hrulefill

\section{Introduction}
\label{sec:intro}

Despite the lack of any direct signal of new physics beyond the Standard Model (SM), the minimal supersymmetric extension of the Standard Model (MSSM) is still one of the forefront candidate because of the motivation of resolution to the gauge hierarchy problem \cite{Gildener:1976ai,Gildener:1979dd,Weinberg:1978ym,Susskind:1978ms,Veltman:1980mj}, stability of the Higgs potential \cite{Degrassi:2012ry,Bezrukov:2012sa,Buttazzo:2013uya,Branchina:2013jra,Branchina:2014usa,R.:2019ply}, pleasant candidates for the dark matter with an additional attraction from the gauge coupling unification at the grand unification scale ($\mgut \simeq 2.4\times 10^{16}$ GeV). On the other hand, the lack of a direct signal might point to deviate from the minimal point of view in constructing models beyond the SM. For instance, if the lightest supersymmetric particle (LSP) is not a mixture of MSSM neutralinos \cite{deGouvea:2006wd,Chala:2017jgg,Potter:2015wsa}, many of the signal processes currently under the collider analyses may not be available at the collision energies of today. A similar discussion can be followed also for the null results from the dark matter experiments \cite{DelleRose:2017uas, DelleRose:2017ukx, Ahmed:2020lua}. 

In addition to the current results from the experiments, possible resolutions to some long standing problems such as absence of the right-handed neutrinos and $\mu -$ problem in MSSM \cite{Bae:2019dgg} can motivate to construct models beyond the SM which extends the particle content and/ or symmetrical structure of MSSM. In this context, a larger symmetry group which supplements the MSSM gauge group with an extra $U(1)'$ can address the resolution to the $\mu-$problem. If the extra $U(1)'$ symmetry is imposed in a way that the MSSM fields are also non-trivially charged under it, 
the $\mu H_{d}H_{u}$ is not allowed in the superpotential due to the gauge invariance under $U(1)'$. On the other hand, it can be generated effectively through the vacuum expectation value (VEV) of a field $S$, which is preferably singlet under the MSSM gauge symmetry so that its VEV breaks only the $U(1)'$ symmetry. In this case the superpotential involves a term 
such as $SH_{u}H_{d}$, and through the $U(1)'$ breaking, the $\mu-$term is generated effectively as $\mu_{{\rm eff}} \sim \mathcal{O}(\langle S \rangle)$ \cite{Cvetic:1997ky,Nelson:2002ca,Ellis:2000js,Frank:2013yta,Bertolini:2010yz,Athron:2009ue,Frank:2020ixv,Suematsu:1994qm,Lee:2007fw,Demir:1998dm}. In this way, the radiative electroweak symmetry breaking (REWSB) is linked to the $U(1)'$ symmetry breaking through the renormalization group equations (RGEs). Such an extension can still be considered to be minimal, and it is well motivated in superstring theories \cite{Cvetic:1996mf}, grand unified theories \cite{Hewett:1988xc} and in dynamical electroweak breaking theories \cite{Hill:2002ap}.

In addition to extending the symmetry, the $U(1)'$ models also extend the 
particle content of MSSM by adding $Z'$ - the gauge boson associated with the $U(1)'$ group and the right handed neutrinos. {The right handed neutrinos can be considered to complete the representations of matter field families so that the model can be embedded in a larger GUT group such as $E_{6}$. Besides, the right-handed neutrinos contribute to the anomaly cancellations, and they can provide a natural framework to implement seesaw mechanisms \cite{Khalil:2010iu} for non-zero neutrino masses and mixing \cite{Wendell:2010md}. In this context}, $\langle S \rangle$ significantly contributes to the right-handed sneutrino and $Z'$ masses as well as their superpartners. {However, the presence of a neutral gauge boson $Z'$ brings a strong impact on this class of models, since the current experimental results exclude the solutions with $M_{Z'} \lesssim 4$ TeV. Such a strong exclusion results in a $U(1)'$ breaking scale at the order of $\mathcal{O}(10~{\rm TeV})$ and consequently $\mu_{eff} \sim {\rm a ~ few~TeV}$, which softly brings the $\mu-$problem back to the $U(1)'$ models, even though a resolution to the naturalness problem can be accommodated \cite{Hicyilmaz:2017nzo}. Note that this strong bound can be avoid if the gauge coupling of $U(1)^{\prime}$ is significantly small ($\lesssim 10^{-5}$) \cite{Okada:2020evk,Nath:2021uqb,Lindner:2020kko}, or negligibly couple to the fermions of the first to families \cite{Abdullah:2019dpu,Frank:2020byg}. However, these conditions on the couplings cannot be met if the gauge coupling unification is imposed at $M_{{\rm GUT}}$ and/or the family universal $U(1)^{\prime}$ charges are assumed. On the other hand, if the $U(1)'$ symmetry breaking involves three more MSSM singlet fields ($S_{1},S_{2},S_{3}$) \cite{Erler:2002pr,Chiang:2008ud} , a $\mu-$term at the order of electroweak symmetry breaking can be realized, while $Z'$ remains heavy to be consistent with the current experimental results. These three MSSM singlet fields form the secluded sector, and their number can be determined by considering the physical properties of vacua of the scalar potential. The minimal setup of $U(1)$ extended supersymmetric models ($\kappa \rightarrow 0$ in $\kappa/3~ S_{1}S_{2}S_{3}$), the unstable vacua of the scalar potentials can be avoided by setting the $U(1)^{\prime}$ breaking scale very large ($\mathcal{O}(10~{\rm TeV})$), which leads to heavy $S$ field and its neutralino. In these cases, the testable low scale implications of $U(1)^{\prime}$ models are not distinguishable from the MSSM implications \cite{Hicyilmaz:2020bph,Frank:2020pui,Hicyilmaz:2017ntm}. Even though extending the scalar sectors with one or two more MSSM singlet fields can allow relatively lower breaking scale for $U(1)^{\prime}$ symmetry, the requirement of physical vacua is still effective, since the solutions in such frameworks can still yield vanishing VEVs for the MSSM Higgs fields or $v_{s}\sim v_{u}\sim v_{d}$, which also cancels the $Z^{\prime}-Z$ hierarchy \cite{Erler:2002pr}. In this context, the minimal form of the secluded sector can be spanned by three MSSM singlet fields. Of course it could include more fields; however, their effects can still be taken into account with three fields by varying the relevant couplings ($\lambda$ and $\kappa$) and VEVs ($v_{S}$ and $v_{i}$).} 

We refer to the class of supersymmetric $U(1)'$ models with three additional MSSM singlet fields as the secluded $U(1)'$ model \cite{Erler:2002pr,Chiang:2008ud,Demir:2010is,Frank:2012ne}. Even though we allow the additional $U(1)^{\prime}$ symmetry can be broken at scale of order multi-TeV ($\mathcal{O}(10)$ TeV), we assume $U(1)-$ extended MSSM emerges at $M_{{\rm GUT}}$ resulting from the breaking of a larger gauge symmetry such as $E_{6}$. We perform random scans by imposing boundary conditions on the GUT scale parameters at $M_{{\rm GUT}}$, which is spanned by the universal soft supersymmetry breaking (SSB) mass terms for the supersymmetric scalars and gauginos. Note that the SSB masses of the Higgs fields are calculated by minimizing the superpotential, and in general, it implies non-universal SSB masses for the MSSM Higgs  fields. Since the breaking scale of $U(1)^{\prime}$ is unknown, the set of the free parameters in secluded $U(1)^{\prime}$ model involves VEVs of the MSSM singlet scalar fields. The rest of the paper is organized as follows: We first briefly review the secluded $U(1)'$ models in Section \ref{sec:model} with its superpotential, particle content, non-trivial $U(1)'$ charges of the fields and anomaly cancellations as well as the field content and the physical mass states. After we describe the scanning procedure and summarize the relevant experimental constraints in Section \ref{sec:scan}, we present and discuss the LHC and  dark matter implications of the model in Section \ref{sec:results}. Finally, we  conclude our findings in Section \ref{sec:conc}.

\section{The Secluded U(1)' Model}
\label{sec:model}
In this section, we present the relevant ingredient and some salient features of the secluded $U(1)'$ model, which is based on the gauge group $SU(3)_{c}\times SU(2)_{L}\times U(1)_{Y}\times U(1)'$. Such an extension of 
MSSM gauge group can emerge in the grand unified theories (GUTs) based on 
a gauge group larger than $SU(5)$ such as $SO(10)$ \cite{DeRujula:1980qc,Derendinger:1983aj,Antoniadis:1987dx,Ellis:1988tx} and/or $E_{6}$ \cite{Lazarides:1986cq,Shafi:1978gg,Gursey:1975ki,Bajc:2013qra}. {In the common convention the $U(1)^{\prime}$ extension of MSSM results from the $E_{6}$ through the following cascade of the symmetry breakings:}

\begin{equation}
\begin{array}{rl}
E_{6} & \rightarrow SO(10)\times U(1)_{\psi} \\
& \rightarrow SU(5)\times U(1)_{\chi}\times U(1)_{\psi} \\
& \rightarrow SU(3)_{C}\times SU(2)_{L}\times U(1)_{Y}\times U(1)^{\prime}
\end{array}
\label{eq:breaking}
\end{equation}
{where $U(1)^{\prime}$ group at the end of the breaking chain is, in principle, a linear combination of $U(1)_{\chi}$ and $U(1)_{\psi}$ with charge $Q^{\prime} = Q_{\chi} \cos \theta_{E_6} + Q_{\psi} \sin \theta_{E_6}$.} On the other hand, a general configuration of the $U(1)'$ charges cannot be restricted to these two classes of the $U(1)'$ models \cite{Bertolini:2010yz,Demir:2005ti,Ellis:2011es,Gogoladze:2009mc,Frank:2020byg,DelleRose:2018eic}. A general set of equations for the charges can be obtained from the anomaly cancellation condition, and these conditions also depend on the exotic fields involved in the model. In our work, we consider the following superpotential:

\begin{equation}
\setstretch{2.0}
\begin{array}{rll}
\widehat{W} & = W_{{\rm MSSM}}(\mu = 0) & + \lambda \widehat{S}\widehat{H}_u \cdot
	\widehat{H}_d + {h_{\nu}} \widehat{L}\cdot
	\widehat{H}_u \widehat{N}+
	\frac{\kappa}{3} \widehat{S}_1 \widehat{S}_2 \widehat{S}_3 \\
	 && +\sum_{i=1}^{n_{\cal{Q}}} {h}_Q^i \widehat{S} \widehat{\cal{Q}}_i
	\widehat{\cal{\overline{Q}}}_i + \sum_{j=1}^{n_{\cal{L}}} {h}_L^j
	\widehat{S} \widehat{\cal{L}}_j \widehat{\cal{\overline{L}}}_j
\end{array}
\label{eq:superpot}
\end{equation}
where the matter superfields of MSSM corresponding to the squarks and sleptons $\widehat{Q}, \widehat{U}, \widehat{D},\widehat{L}$ and $\widehat{E}$ are included in $W_{{\rm MSSM}}$, and $\widehat{H}_{u}, \widehat{H}_{d}$ are the MSSM Higgs doublets. The new ingredient from the $U(1)'$ inclusion can be listed as the MSSM singlet scalars $S,S_{1,2,3}$, right-handed neutrino superfield $\widehat{N}$ and exotic fields $\mathcal{Q}_{i}$, $\mathcal{L}_{i}$. In addition, the model includes a neutral gauge boson associated with the $U(1)'$ symmetry and its supersymmetric partner. {Note that we do not assume $Q_{H_u}^{\prime} + Q_{H_d}^{\prime} = 0$ in our scans. In this case, the gauge invariance forbids the bilinear mixing ($\mu \widehat{H}_{u}\widehat{H}_{d}$) in the MSSM superpotential, which is stated as $W_{{\rm MSSM}}(\mu=0)$ in Eq.(\ref{eq:superpot}).} It is rather generated effectively through the VEV of $S$ so that $\mu_{eff} \equiv \lambda \langle S \rangle$. However, emergence of such an effective term induces mixed anomalies between $U(1)'$ and the MSSM gauge group, and cancellation of such anomalies also requires exotic fields in the particle spectrum, and the anomaly cancellation can be maintained by introducing exotic fields, which are vector-like with respect to MSSM, but chiral under the $U(1)'$ group. 

\begin{table}[h]
\centering
\scalebox{0.8}{
	\begin{tabular*}{1.3\textwidth}{@{\extracolsep{\fill}} ccccccccccccccccc}
		\hline \hline Field & $\widehat{Q}$ & $\widehat{U}$ &
		$\widehat{D}$ & $\widehat{L}$ & $\widehat{N}$ & $\widehat{E}$ &
		$\widehat{H}_u$ & $\widehat{H}_d$ & $\widehat{S}$ & $\widehat{S_1}$
		& $\widehat{S_2}$ & $\widehat{S_3}$ & $\widehat{\cal{Q}}$ &
		$\widehat{\cal{\overline{Q}}}$ & $\widehat{\cal{L}}$ &
		$\widehat{\cal{\overline{L}}}$
		\\\hline
		$\;$  $SU(3)_C$ & 3 & $\overline{3}$ &  $\overline{3}$ & 1 & 1 &  1& 1& 
1 &1
		& 1& 1 &1 &3 &$\overline{3}$
		&1&1\\
		$\;$  $SU(2)_L$ & 2 & 1 &  1 & 2 & 1 &  1& 2& 2 &1 & 1& 1 &1 &1 &1
		&1 &1\\
		$\;$  $U(1)_Y$ & 1/6 & -2/3 &  1/3 & -1/2 & 0 &  1&
		1/2& -1/2 &0 & 0& 0 &0 &$Y_{\cal{Q}}$ &$-Y_{\cal{Q}}$ &$Y_{\cal{L}}$
		& $-Y_{\cal{L}}$\\
		$\;$  $U(1)^{\prime}$ & $Q_{Q}^{\prime}$ &
		$Q_{U}^{\prime}$ & $Q_{D}^{\prime}$& $Q_{L}^{\prime}$ &
		$Q_{N}^{\prime}$& $Q_{E}^{\prime}$& $Q_{H_u}^{\prime}$&
		$Q_{H_d}^{\prime}$& $Q_{S}^{\prime}$ & $Q_{S_1}^{\prime}$&
		$Q_{S_2}^{\prime}$ &$Q_{S_3}^{\prime}$ &$Q_{\cal{Q}}^{\prime}$
		&$Q_{\cal{\overline{Q}}}^{\prime}$ &$Q_{\cal{L}}^{\prime}$
		&$Q_{\cal{\overline{L }}}^{\prime}$
		\\\hline\hline
	\end{tabular*}}
	\caption{\sl\small Gauge quantum numbers of quark ($\widehat{Q},
		\widehat{U}, \widehat{D}$), lepton ($\widehat{L}, \widehat{N},
		\widehat{E}$), Higgs ($\widehat{H}_u, \widehat{H}_d$), MSSM-singlet
		($\widehat{S}, \widehat{S}_1, \widehat{S}_2, \widehat{S}_3$), exotic
		quark ($\widehat{\cal{Q}}, \widehat{\cal{\overline{Q}}}$) and exotic
		lepton ($\widehat{\cal{L}}, \widehat{\cal{\overline{L}}}$)
		superfields.} \label{tab:charge}
\end{table}

If a general $U(1)'$ charge assignments as shown in Table $\ref{tab:charge}$, the gauge invariance condition yields the following equations:

\begin{eqnarray}
	\label{eq:gauge_cond}
	0&=&Q^{\prime}_{S}+Q^{\prime}_{H_u}+Q^{\prime}_{H_d}~,
	\nonumber \\
	0&=&Q^{\prime}_{Q}+Q^{\prime}_{H_u}+Q^{\prime}_{U}~,
	\nonumber \\
	0&=&Q^{\prime}_{Q}+Q^{\prime}_{H_d}+Q^{\prime}_{D}~,
	\nonumber \\
	0&=&Q^{\prime}_{L}+Q^{\prime}_{H_d}+Q^{\prime}_{E}~,
	\nonumber \\
	0&=&Q^{\prime}_{\cal{Q}}+Q^{\prime}_{\cal{\overline{Q}}}+Q^{\prime}_{S}~,
	\nonumber \\
	0&=&Q^{\prime}_{\cal{L}}+Q^{\prime}_{\cal{\overline{L}}}+Q^{\prime}_{S}~,
	\nonumber \\
	0&=&Q^{\prime}_{L}+Q^{\prime}_{H_u}+Q^{\prime}_{N}~,
	\nonumber \\
	0&=&Q^{\prime}_{S_1}+Q^{\prime}_{S_2}+Q^{\prime}_{S_3}~.
\end{eqnarray}
Note that if a special configuration with $Q'_{S}=0$ can be found, the $\mu-$term becomes allowed by the gauge invariance. However, a consistent $Z-Z'$ mass hierarchy and mixing rather require non-zero $U(1)'$ charges for 
all MSSM singlet scalar fields $S,S_{1,2,3}$. Another set of conditions for the charges is obtained from the vanishing $U(1)'-SU(3)_{c}-SU(3)_{c}$, $U(1)'-SU(2)_{L}-SU(2)_{L}$, $U(1)'-U(1)_{Y}-U(1)_{Y}$, $U(1)'-{\rm graviton-graviton}$, $U(1)'-U(1)'-U(1)_{Y}$ and $U(1)'-U(1)'-U(1)'$ anomalies, as follows:

\begin{eqnarray}
    \label{eq:anomaly_cond}
	0&=&3(2Q^{\prime}_{Q}+Q^{\prime}_{U}+Q^{\prime}_{D})+
	n_{\cal{Q}}(Q^{\prime}_{\cal{Q}}+Q^{\prime}_{\cal{\overline{Q}}})~,
	\\
	0&=&3(3Q^{\prime}_{Q}+Q^{\prime}_{L})+Q^{\prime}_{H_d}+Q^{\prime}_{H_u}~,
	\\
	0&=&3(\frac{1}{6}Q^{\prime}_{Q}+\frac{1}{3}Q^{\prime}_{D}+
	\frac{4}{3}Q^{\prime}_{U}+
	\frac{1}{2}Q^{\prime}_{L}+Q^{\prime}_{E})
	+\frac{1}{2}(Q^{\prime}_{H_d}+Q^{\prime}_{H_u})\nonumber \\
	&+&3n_{\cal{Q}} Y^2_{\cal{Q}} (Q^{\prime}_{\cal{Q}}+
	Q^{\prime}_{\cal{\overline{Q}}})+ n_{\cal{L}}
	Y^2_{\cal{L}} (Q^{\prime}_{\cal{L}}+
	Q^{\prime}_{\cal{\overline{L}}})~,
	\\
	0&=&3(6Q^{\prime}_{Q}+3Q^{\prime}_{U}+3Q^{\prime}_{D}+2Q^{\prime}_{L}+
	Q^{\prime}_{E}+Q^{\prime}_{N})
	+2Q^{\prime}_{H_d}+2Q^{\prime}_{H_u}\nonumber \\
	&+&Q^{\prime}_{S}+Q^{\prime}_{S_1}+Q^{\prime}_{S_2}+Q^{\prime}_{S_3}+
	3 n_{\cal{Q}}
	(Q^{\prime}_{\cal{Q}}+Q^{\prime}_{\cal{\overline{Q}}})+
	n_{\cal{L}}(Q^{\prime}_{\cal{L}}+Q^{\prime}_{\cal{\overline{L}}})~,
	\\
	0&=&3(Q^{\prime\ 2}_{Q}+Q'^2_{D}-2Q^{\prime\
		2}_{U}-Q^{\prime\ 2}_{L}+Q^{\prime\ 2}_{E})-Q^{\prime\
		2}_{H_d}+Q^{\prime\ 2}_{H_u}+ 3n_{\cal{Q}} Y_{\cal{Q}}
	(Q^{\prime\ 2}_{\cal{Q}}- Q^{\prime\
		2}_{\cal{\overline{Q}}})\nonumber \\
	&+&n_{\cal{L}} Y_{\cal{L}}
	(Q^{\prime\ 2}_{\cal{L}}-Q^{\prime\
		2}_{\cal{\overline{L}}})~,
	\\
	0&=&3(6Q^{\prime\ 3}_{Q}+3Q^{\prime\ 3}_{D}+3Q^{\prime\
		3}_{U}+2Q^{\prime\ 3}_{L}+Q^{\prime\ 3}_{E}+Q^{\prime\
		3}_{N})+ 2Q^{\prime\ 3}_{H_d}+2Q^{\prime\
		3}_{H_u}+Q^{\prime\ 3}_{S}\nonumber \\
	&+&Q^{\prime\
		3}_{S_1}+Q^{\prime\ 3}_{S_2}+Q^{\prime\ 3}_{S_3}+
	3n_{\cal{Q}}(Q^{\prime\ 3}_{\cal{Q}}+Q^{\prime\
		3}_{\cal{\overline{Q}}})+n_{\cal{L}}(Q^{\prime\
		3}_{\cal{L}}+Q^{\prime\ 3}_{\cal{\overline{L}}})~.
	\label{eq:anomaly_cond2}
\end{eqnarray}

All these conditions from the gauge invariance and the anomaly cancellations should be satisfied for particular pattern of charges and parameters, which requires the number of exotics. {Based on the choice of exotics in the model, one of the simplest solution to the mixed anomaly constraints requires  $n_{\mathcal{Q}}=3$ color triplets with $Y_{\mathcal{Q}}=-1/3$ and $n_{\mathcal{L}}=2$ color singlets with $Y_{\mathcal{L}}=-1$. Recall that these exotic fields are singlets under $SU(2)_{L}$ as mentioned before and listed in Table \ref{tab:charge}. In our analyses, we assume the exotics to be very heavy and decouple from the low scale spectrum. In this sense, their charges are considered only for the anomaly cancellation in our work.}

\subsection{Gauge Boson Masses and Mixing}

As mentioned before the model introduces a new neutral gauge boson $Z'$ and its superpartner $\tilde{B}'$ associated with the gauged $U(1)'$ symmetry. The 
symmetry breaking in this model is being realized very similar to the Higgs mechanism, but in this case, the electroweak and $U(1)'$ symmetry breaking are correlated. The fields developing non-zero VEVs during the symmetry breaking, $SU(2)_{L}\times U(1)_{Y}\times U(1)'\rightarrow U(1)_{{\rm EM}}$, can be listed as follows:

\begin{equation}
\langle H_{u} \rangle = \dfrac{1}{\sqrt{2}}\left(\begin{array}{c} 0 \\ v_{u} \end{array}\right)~,\hspace{0.3cm}
\langle H_{d} \rangle = \dfrac{1}{\sqrt{2}}\left(\begin{array}{c} 0 \\ v_{d} \end{array}\right)~,\hspace{0.3cm}
\langle S \rangle = \dfrac{v_{S}}{\sqrt{2}}~,\hspace{0.3cm}
\langle S_{i} \rangle = \dfrac{v_{S_{i}}}{\sqrt{2}}
\label{eq:vevs}
\end{equation}

Since $S,S_{1,2,3}$ fields are singlet under the MSSM gauge group, the $W$ and $Z-$bosons acquire their masses through the VEVs of $H_{u}$ and $H_{d}$ as in the usual electroweak symmetry breaking; thus, the condition $v_{u}^{2}+v_{d}^{2}=v_{SM}^{2}$ should still hold in this model. On the other hand, {because of the non-trivial charges of all the superfields under $U(1)'$ as listed in Table \ref{tab:charge}}, the $Z'-$boson receives its mass from all the VEVs. In this case, since $v_{S}$ and/or $v_{S_{i}}$ are expected to be much greater than $v_{u}$ and $v_{d}$, the secluded sector can be 
accounted for the main source of the $Z'$ mass. However, apart from the mass acquisition, non-trivial $H_{u}$ and $H_{d}$ charges induce 
a non-zero mixing between $Z$ and $Z'$ associated with their mass-square matrix:

\begin{equation}
\setstretch{2}
M^{2}_{ZZ'} = \left( 
\begin{array}{ccc}
M_{Z}^{2} & &\Delta^{2} \\ \Delta^{2} & & M^{2}_{Z'}
\end{array}\right)
\label{eq:mZZp}
\end{equation}
written in $(Z,Z')$ basis in terms of 
\begin{equation}
\setstretch{2}
\begin{array}{ll}
M_{Z}^{2}& = \dfrac{1}{4}(g_{1}^{2}+g_{2}^{2})(v_{u}^{2}+v_{d}^{2}) \\
M^{2}_{Z'} & = g^{'2}_{1}\left( Q^{'2}_{H_{u}}v_{u}^{2}+Q^{'2}_{H_{d}}v_{d}^{2}+Q^{'2}_{S}v_{S}^{2}+\displaystyle \sum_{i=1}^{3}Q^{'2}_{S_{i}}v_{S_{i}}^{2} \right) \\
\Delta^{2} & = \dfrac{1}{2}\sqrt{g_{1}^{2}+g_{2}^{2}}~g'_{1}(Q^{'2}_{H_{u}}v_{u}^{2}-Q^{'2}_{H_{d}}v_{d}^{2})
\end{array} 
\end{equation}

Diagonalizing the mass-square matrix in Eq.(\ref{eq:mZZp}) yields the following mixing angle between $Z$ and $Z'$:

\begin{equation}
\theta_{ZZ'} = \dfrac{1}{2}\arctan\left( \dfrac{2\Delta^{2}}{M_{Z'}^{2}-M_{Z}^{2}}\right)
\end{equation}
and the electroweak precision data strongly bounds the $Z-Z'$ mixing angle as $\theta_{ZZ'} \lesssim 10^{-3}$ \cite{Erler:2009jh,delAguila:2010mx,Aaltonen:2010ws}. Applying such a strict constraint to the mixing between 
$Z$ and $Z'$ allows only solutions {with} the following properties:

\begin{enumerate}
\item $g'_{1} \ll g_{1}$, or
\item $M_{Z'} \gg M_{Z}$, or
\item $Q'_{H_{d}}/Q'_{H_{u}} \simeq v_{u}/v_{d}\equiv \tan\beta$.
\end{enumerate}
The first two conditions separately bring a naive application of the LEP 2 bound on $M_{Z'}$ as $M_{Z'}/g'_{1} \geq 6$ TeV \cite{Cacciapaglia:2006pk}. The first condition cannot be realized when the secluded model is constrained by the gauge coupling unification at $\mgut$, since the gauge coupling unification yields $g'_{1} \sim g_{1},g_{2}$ at the low scale. Thus, a consistent $Z-Z'$ mixing with the precision data can be satisfied by spectra involving heavy $Z'$. Alternatively, one can apply the third condition by adjusting $Q'_{H_{u}}$ and $Q'_{H_{d}}$; however, since it enhances the model dependency in 
the results, we do not consider it in our study.

{Note that the gauge invariance also allows a tree-level mixing between the gauge bosons of two abelian gauge groups through $\xi B^{\mu\nu}B^{\prime}_{\mu\nu}$, where $B_{\mu\nu}$ and $B^{\prime}_{\mu\nu}$ correspond to the field strength tensors for $~U(1)_{Y}$ and $U(1)^{\prime}$ in our work respectively. This term leads to a gauge covariant derivative in a non-canonical form \cite{OLeary:2011vlq,Un:2016hji}, which also induces tree-level mixing between the MSSM Higgs fields and MSSM singlet scalar fields of the secluded sector. In addition, it also induces a mixing between the photon and $Z^{\prime}$ in addition to $Z-Z^{\prime}$ mixing, which alters the mixing angle given in Eq.(2.13) such that the physical mass states of the gauge bosons involve a massless photon \cite{Langacker:2008yv,Babu:1997st}. Even though the gauge kinetic mixing can yield relatively lighter $Z^{\prime}$ in the mass spectrum \cite{Langacker:1998tc}, it is severely bounded from above. The analyses over the LEP data yield an upper bound on the gauge kinetic mixing as $\xi \lesssim 2.5\times 10^{-3}$ \cite{Erler:2009jh}, and the current collider analyses have upgraded this bound as $\xi \lesssim 3\times 10^{-4}$ from searches over different decay modes of $Z^{\prime}$ \cite{Pankov:2019yzr,Bobovnikov:2018fwt,CMS:2016wev}. In addition, the direct detection experiments can severely restrict the gauge kinetic mixing through the scattering of DM, since it raises the photon and $Z-$boson abundance \cite{Lao:2020inc}, unless the scattering is not sufficiently suppressed by the $Z^{\prime}$ mass. Thus, the gauge kinetic mixing does not loose the severe bound on the $Z^{\prime}$ mass which is also required by the resonance searches of the collider analyses \cite{ATLAS:2019erb,CMS:2019tbu}. In this context, we assume the available spectra in our model involve heavy $Z^{\prime}$, and we set the gauge kinetic mixing to zero, since its impact would be negligible after imposing the relevant constraints on it.}

\subsection{Neutralinos and Charginos}

Depending on the hypercharges of the exotic fields ($\mathcal{Q}$ and $\mathcal{L}$), the secluded $U(1)'$ model extends both the charged and neutral sectors of MSSM. Their interference enriches the phenomenology such as lowering the mass bound on $Z'$ \cite{Langacker:1998tc,Kang:2004bz}, triggering $U(1)'$ breaking by guaranteeing negative $m_{S}^{2}$ \citep{Cvetic:1997ky} etc. Even though it is possible to have exotic fields coupling the quarks to leptons depending on baryon and lepton numbers \cite{Lazarides:1998iq}, in the standard configuration they can couple only to quarks or leptons. Besides, it is possible to configure the $U(1)'$ charges in which the exotics are allowed to couple only to $S$ field at tree level. In this case, $SU(3)$ triplet exotic field $\mathcal{Q}$ is still allowed to be produced at the collider experiments; thus the resonance searches are still able to bound their masses as $m_{\tilde{Q}} \gtrsim 5$ TeV \cite{Kaur:2020fja}. In this context, we assume the exotics can couple only to the MSSM singlet $S$ field, and they are heavy. Thus their observable effects at the low scale become suppressed, while they are still effective in 
$U(1)'$ symmetry breaking, anomaly cancellation etc.

Assuming the exotic fields to be decoupled at a high scale only the {MSSM singlet} fields can be involved in the spectrum, and the neutral sector of this class of secluded $U(1)'$ models significantly extends the Neutralinos with $\tilde{S}, \tilde{B}', \tilde{S}_{1,2,3}$. After the 
$U(1)'$ and electroweak symmetry breakings, these neutralinos together with the MSSM neutralinos mix each other, and the resultant mass matrix for 
the neutralino sector can be obtained {in the usual basis ordered as ($\tilde{B}$, $\tilde{W}$, $\tilde{H}_{d}$, $\tilde{H}_{u}$, $\tilde{S}$, $\tilde{B}^{\prime}$, $\tilde{S}_{1}$, $\tilde{S}_{2}$, $\tilde{S}_{3}$)} as follows:

\begin{equation}
\setstretch{1.5}
\resizebox{0.85\hsize}{!}{$
M_{\tilde{\chi}}=\left(\begin{array}{ccccc|ccccc}
&&&&& 0 & M_{BB'} & 0 & 0 & 0 \\
&&&& &0 & 0 & 0 & 0 &0 \\
\multicolumn{4}{c}{{\rm MSSM(\mu=\mu_{{\rm eff}})}} & &- \dfrac{\lambda 
v_{u}}{2} & g'_{1}Q'_{H_{d}}v_{d} & 0 & 0 & 0 \\
&&&&& - \dfrac{\lambda v_{d}}{2} & g'_{1}Q'_{H_{u}}v_{u} & 0 & 0 & 0 \\ &&&&&&&& \\ \hline &&&&&&&& \\
0 & 0 & -\dfrac{\lambda v_{u}}{2} & -\dfrac{\lambda v_{d}}{2} & & 0 & g'_{1}Q'_{S}v_{S} & 0 & 0 & 0 \\
M_{BB'} & 0 & g'_{1}Q'_{H_{d}}v_{d} & g'_{1}Q'_{H_{u}}v_{u} & & g'_{1}Q'_{S}v_{S}  & M_{B'} &  g'_{1}Q'_{S_{1}}v_{S_{1}} & g'_{1}Q'_{S_{2}}v_{S_{2}} & g'_{1}Q'_{S_{3}}v_{S_{3}} \\
0 & 0 & 0 & 0 & & 0 & g'_{1}Q'_{S_{1}}v_{S_{1}} & 0 & -\dfrac{\kappa v_{S_{3}}}{\sqrt{2}} & -\dfrac{\kappa v_{S_{2}}}{\sqrt{2}} \\
0 & 0 & 0 & 0 & & 0 & g'_{1}Q'_{S_{2}}v_{S_{2}} & -\dfrac{\kappa v_{S_{3}}}{\sqrt{2}} & 0 & -\dfrac{\kappa v_{S_{1}}}{\sqrt{2}} \\
0 & 0 & 0 & 0 & & 0 & g'_{1}Q'_{S_{2}}v_{S_{3}} & -\dfrac{\kappa v_{S_{2}}}{\sqrt{2}} & -\dfrac{\kappa v_{S_{1}}}{\sqrt{2}} & 0
\end{array} \right)$}
\label{eq:Mneutralino}
\end{equation}

The upper block called MSSM in the mass matrix given above represents the  usual MSSM neutralinos and their mixing. However, since the $\mu-$term is 
not allowed at tree level and effectively generated by the VEV of $S$, the secluded $U(1)'$ is effective in generating the masses of MSSM Higgsinos and interfering in mixing of MSSM neutralinos. {In addition, $\tilde{B}^{\prime}$ can mix with $\tilde{B}$ through the kinetic mixing, and MSSM Higgsinos through $Z-Z^{\prime}$ mixing which are quantified with $M_{BB^{\prime}}$, $g'_{1}Q'_{H_{d}}v_{d}$ and $g'_{1}Q'_{H_{u}}v_{u}$ in the neutralino mass matrix, respectively. Similarly, $S$ can mix with the MSSM Higgsinos and its mixing is proportional to its coupling to the Higgs fields as $\lambda v_{d}$ and $\lambda v_{u}$. On the other hand, $\tilde{S}_{1}$, $\tilde{S}_{2}$ and $\tilde{S}_{3}$ mix only with the $U(1)^{\prime}$ neutralinos, while they leave the MSSM neutralinos intact.} 

In addition to their effects in the neutralino sector, these fields including $\tilde{B}'$ can also escape from the experimental detection and they can easily be much lighter than the MSSM neutralinos. For instance, heavy mass bounds on gluino as $m_{\tilde{g}}\ge 2.1$ TeV \cite{Aaboud:2017vwy} also bounds the Bino and Wino masses at about 300 GeV and 600 GeV, respectively when the universal gaugino mass is imposed at the GUT scale (for a recent study with universal gaugino mass at the GUT scale, see \cite{Babu:2020ncc}). In addition, the current measurements of the Planck satellite on the relic density of the dark matter can lift the mass bound up to about 1 TeV, especially when the dark matter is composed mostly by Bino \cite{Ahmed:2020lua}. {Even though the LHC bounds can be loosen when the LSP is formed mostly by the MSSM Higgsinos ($\mu \ll M_{1},M_{2}$), the current null results from the direct detection experiments require $\mu\gtrsim 700$ GeV for the Higgsino-like dark matter \cite{Ahmed:2020lua,Raza:2018jnh,Gomez:2020gav}. Note that this bound is significantly  reduced to about $200-300$ GeV if the DM is realized to be Higino-Bino and/or Higgsino-Wino mixture \cite{Krall:2017xij}.}


As a consequence of such severe bounds on the MSSM neutralinos, the non-MSSM neutralinos can be more likely to form the LSP neutralino in the low scale mass spectrum, and in this context they yield quite different phenomenology in both the collider and the dark matter experiments. If the LSP 
is formed mostly by the secluded $U(1)'$ sector, then some of the particles in the MSSM spectrum might be realized to be a long lived state, since 
they do not directly couple to the LSP. Even though, the current LHC constraints on the strongly interacting particles such as squarks and gluino yield consistent lifetime for these particles, it is still possible to have long lived staus and charginos in the low scale spectrum, and the model should 
be constrained to avoid possible missing electric charges from such states escaping from detectors. 

On the other hand, even though the stop and gluino are not allowed to be long lived by the LHC constraints, the current bounds on these particles can be significantly modified depending on the decay of the lightest MSSM neutralino into LSP. Since these particles do not directly couple to the LSP, their possible signal processes remain the same as those which are excessively analyzed in the collider experiments \cite{Vami:2019slp}. In these processes, if the lightest MSSM neutralino does not decay in the detector and 
it forms the missing energy, then the current constraints on the stop and 
gluino still hold. On the other hand, if the lightest MSSM neutralino is allowed to decay into LSP in the detector, then such processes can significantly modify (probably loosen) the current bounds on the stop and gluino (see \cite{Chala:2017jgg} for the case in which stop does not directly couple to LSP). 

The non-MSSM LSP also yields an interesting phenomenology in the dark matter experiments. Since $S$ is allowed to interact with the MSSM Higgs fields at tree-level, its superpartner ($\tilde{S}$) scatters at nuclei through Higgs portal. Even though its scattering cross-section is expected to 
be rather low, such solutions will be able to be tested soon under the current 
and future projected sensitivity of the results from XENON experiment \cite{Aprile:2020vtw}.  The scattering cross-section can be further lowered when the LSP is formed by $\tilde{S}_{1,2,3}$, since they interact only with $S$. However, their annihilation processes can still yield interesting results for the indirect detection of dark matter and can be tested under light of the current results from FermiLAT \cite{Ackermann:2015zua,Drlica-Wagner:2015xua}. 

Before concluding the neutralino sector in the model, we should also note 
the chargino sector. Since only the {MSSM singlet} fields can be involved in the 
detectable low scale spectrum, the physical states of the chargino sector 
is formed by the MSSM fields such as Wino and Higgsino. On the other hand, as discussed in the neutralino mass matrix, the charged Higgsino mass ($\mu$) is determined effectively by the VEV of $S$, thus the secluded $U(1)'$ sector is still effective in the phenomenology of the chargino sector.

\subsection{Higgs Bosons}

The presence of the MSSM singlet $S$ and $S_{i}$ fields significantly extends the Higgs boson sector of MSSM in the secluded $U(1)'$ model. In the physical spectrum there are six CP-even Higgs bosons, while the number of the CP-odd Higgs bosons is four. In addition to the enrichment in the Higgs 
boson spectrum, the Higgs sector becomes more complicated through the tree-level mixing, which does not exist in the MSSM framework. The Higgs potential generated with $F-$terms  can be written as \cite{Chiang:2008ud}

\begin{equation}
V_{F}=|\lambda|^{2}\left[|H_{d}H_{u}|^{2} +|S|^{2}(H_{d}^{\dagger}H_{d}+H_{u}^{\dagger}H_{u})\right]+\dfrac{|\kappa|^{2}}{9}\left(|S_{1}S_{2}|^{2}+|S_{2}S_{3}|^{2}+|S_{1}S_{3}|^{2} \right]
\label{eq:VF}
\end{equation}
where the $SU(2)_{L}$ indices are suppressed for simplicity. While the F-terms allow mixing only among the MSSM Higgs doublets and the MSSM singlet $S$ scalar, the other MSSM singlets can mix with the MSSM Higgs fields through the scalar potential generated by $D-$terms, which is

\begin{equation}
\setstretch{2}
\begin{array}{ll}
V_{D} = & \dfrac{g_{1}^{2}+g_{2}^{2}}{8}\left(H_{d}^{\dagger}H_{d}-H_{u}^{\dagger}H_{u} \right)^2+ \dfrac{g_{2}^{2}}{2}|H_{d}^{\dagger}H_{u}|^{2} \\
 & +\dfrac{g'^{2}_{1}}{2}\left(Q'_{H_{d}}H_{d}^{\dagger}H_{d}+Q'_{H_{u}}H_{u}^{\dagger}H_{u}+Q'_{S}|S|^{2}+\displaystyle \sum_{i=1}^{3}Q'_{S}|S_{i}|^{2} 
 \right)^{2} 
\end{array}
\label{eq:VD}
\end{equation}

After all, the scalar potential generated by the $F-$ and $D-$terms induce tree-level mixing among all the Higgs scalars. In addition, the soft supersymmetry breaking (SSB) Lagrangian contribute to the Higgs phenomenology through the following terms:

\begin{equation}
\setstretch{2}
\begin{array}{ll}
\mathcal{L}_{\cancel{{\rm SUSY}}} = & m_{H_{d}}^{2}H_{d}^{\dagger}H_{d} 
+ m_{H_{u}}^{2}H_{u}^{\dagger}H_{u}+m_{S}^{2}|S|^{2}+\displaystyle \sum_{i=1}^{3}m_{S_{i}}^{2}|S_{i}|^{2} \\
 & -\lambda A_{\lambda} SH_{d}H_{u} - \dfrac{\kappa}{3}A_{\kappa}S_{1}S_{2}S_{3}-m_{SS_{1}}^{2}SS_{1}-m_{SS_{2}}^{2}SS_{2}-m_{S_{1}S_{2}}^{2}S^{\dagger}_{1}S_{2}
\end{array}
\label{eq:Soft}
\end{equation}

The tree-level mass-square matrix for CP-even and CP-odd Higgs bosons is generated through the SSB masses and the vacuum expectation values (VEVs) defined as

\begin{equation*}
H_{d}=\dfrac{1}{\sqrt{2}}\left( \begin{array}{cc}
v_{d}+h^{0}_{d}+iA_{d} \\ \sqrt{2}h_{d}^{-}
\end{array}  \right), \hspace{0.3in}
H_{u}=\dfrac{1}{\sqrt{2}}\left( \begin{array}{cc}
\sqrt{2}h_{u}^{+} \\ v_{u}+h^{0}_{u}+iA_{u}
\end{array}  \right) \tag{2.17-a}
\label{eq:MSSMVEV}
\end{equation*}
\begin{equation*}
S = \dfrac{1}{\sqrt{2}}(v_{S}+S+iA_{S}),\hspace{0.3in} S_{i} = \dfrac{1}{\sqrt{2}}(v_{S_{i}}+S_{i}+iA_{S_{i}}) \tag{2.17-b}
\label{eq:SingletVEV}
\end{equation*}
\setcounter{equation}{17}

Note that the gauge boson associated with $U(1)'$ group ($Z'$) receives its mass from the VEVs of all the scalar fields given in Eqs.(\ref{eq:MSSMVEV}, \ref{eq:SingletVEV}); however, since $v_{S},v_{S_{i}} \gg v_{d},v_{u}$, the VEVs of MSSM singlet fields are dominant in $Z'$ mass. Thus, a heavy mass bound on $Z'$ is expected to yield a strong impact in the MSSM singlet scalar sector. In a class of $U(1)'$ extended SUSY models, the absence of the fields $S_{i}$ results in high $U(1)'$ breaking scale ($v_{S} \gtrsim 10$ TeV) \cite{Hicyilmaz:2017nzo,Hicyilmaz:2016kty}  to realize $M_{Z'} \geq 4-5$ TeV  \cite{Aad:2019fac}. In addition, the VEVs, in principle do not have to align in the same direction, so they can be also a source for CP-violation. However, we assume the CP-conservation in our work by setting $\theta_{i}=0$ (for a detailed discussion about CP-conservation and breaking, see \cite{Chiang:2008ud}).  

The scalar potentials involving the Higgs fields yield the following symmetric mass-square matrix for the CP-even Higgs fields:

\begin{equation}
\mathcal{M}^2_{{\rm even}} = \left( 
\begin{array}{cccccc}
M_{11}^{2} & M_{12}^{2} & M_{13}^{2} & M_{14}^{2} & M_{15}^{2} & M_{16}^{2} \\
 & M_{22}^{2} & M_{23}^{2} & M_{24}^{2} & M_{25}^{2} & M_{26}^{2} \\
 & & M_{33}^{2} & M_{34}^{2} & M_{35}^{2} & M_{36}^{2} \\
& &  & M_{44}^{2} & M_{45}^{2} & M_{46}^{2} \\
 &  &  &  & M_{55}^{2} & M_{56}^{2} \\
 & &  &  &  & M_{66}^{2} \\
\end{array}
\right)
\end{equation}
where, in the basis of $\lbrace  h_{d}, h_{u}, S, S_{1}, S_{2}, S_{3}    \rbrace$,

\begin{equation*}
\setstretch{2}
\Scale[0.75]{
\begin{array}{l}
M_{11}^{2} = \dfrac{(g_{1}^{2}+g_{2}^{2})v_{d}^{2}}{4}+g'^{2}_{1}Q_{H_{d}}^{\prime2}v_{d}^{2}+\dfrac{A_{\lambda}\lambda v_{S}v_{u}}{\sqrt{2}v_{d}} \\
M_{12}^{2}=-\dfrac{A_{\lambda}\lambda v_{S}}{\sqrt{2}}- \dfrac{(g_{1}^{2}+g_{2}^{2})v_{d}v_{u}}{4}+\left(\lambda^{2} + g'^{2}_{1}Q'_{H_{d}}Q'_{H_{u}} \right)v_{d}v_{u} \\
M_{13}^{2}=\lambda^{2}v_{d}v_{s}+g'^{2}_{1}Q'_{H_{d}}Q'_{S}v_{d}v_{S}-\dfrac{A_{\lambda}\lambda v_{u}}{\sqrt{2}} \\
M_{1i+3}^{2}=g'^{2}_{1}Q'_{H_{d}}Q'_{S_{i}}v_{d}v_{S_{i}}~, \hspace{0.3in} i=1,2,3 \\
M_{22}^{2}=\dfrac{A_{\lambda}\lambda v_{d}v_{S}}{\sqrt{2}v_{u}}+\dfrac{1}{4}\left(g_{1}^{2}+g_{2}^{2} + 4g'^{2}_{1}Q'_{H_{u}}\right) v_{u}^{2} \\
M_{23}^{2}=-\dfrac{A_{\lambda}\lambda v_{d}}{\sqrt{2}} + \left(\lambda^{2}+g'^{2}_{1}Q'_{H_{u}}Q'_{S}\right)v_{u}v_{S}\\
M_{2i+3}^{2}=g'^{2}_{1}Q'_{H_{u}}Q'_{S_{i}}v_{u}v_{S_{i}}~, \hspace{0.3in} i=1,2,3
\end{array}\hspace{0.5cm}
\begin{array}{l}
M_{33}^{2} = \dfrac{1}{2v_{S}}\left(2g'^{2}_{1}Q_{S}^{\prime2}v_{s}^{3} -2m_{SS_{1}}^{2}v_{S_{1}}- 2m_{SS_{2}}^{2}v_{S_{2}}+\sqrt{2}A_{\lambda}\lambda v_{d}v_{u}\right) \\
M_{3i+3}^{2}=m_{SS_{i}}^{2}+g'^{2}_{1}Q'_{S}Q'_{S_{i}}v_{S}v_{S_{i}}~\hspace{0.3in} i=1,2,3~{\rm and}~m_{SS_{3}}=0 \\
M_{44}^{2} = \dfrac{1}{2v_{S_{1}}}\left(2g'^{2}_{1}Q_{S_{1}}^{\prime2}v_{S_{1}}^{3}-2m_{SS_{1}}^{2}v_{S}+\sqrt{2}A_{\kappa}\kappa v_{S_{2}}v_{S_{3}} \right) \\
M_{45}^{2} = \dfrac{1}{9}\kappa^{2}v_{S_{1}}v_{S_{2}}+g'^{2}_{1}Q'_{S_{1}}Q'_{S_{2}}v_{S_{1}}v_{S_{2}}-\dfrac{A_{\kappa}\kappa v_{S_{3}}}{\sqrt{2}} \\
M_{46}^{2} = \dfrac{1}{9}\left(\kappa^{2}+9g'^{2}_{1}Q'_{S_{1}}Q'_{S_{3}} \right) v_{S_{1}}v_{S_{3}}   - \dfrac{A_{\kappa}\kappa v_{S_{2}}}{\sqrt{2}} \\
M_{55}^{2}=\dfrac{1}{2v_{S_{2}}}\left(2g'^{2}_{1}Q_{S_{2}}^{\prime2}v_{S_{2}}^{3}-2m_{SS_{2}}^{2}v_{S}+\sqrt{2}A_{\kappa}\kappa v_{S_{1}}v_{S_{3}}\right) \\
M_{56}^{2}=\dfrac{1}{9}\left( \kappa^{2}+9g'^{2}_{1}Q'_{S_{2}}Q'_{S_{3}}\right) v_{S_{2}}v_{S_{3}}-\dfrac{A_{\kappa}\kappa v_{S_{1}}}{\sqrt{2}} \\
M_{66}^{2} = g'^{2}_{1}Q_{S_{3}}^{\prime2}v_{S_{3}}^{2} + \dfrac{A_{\kappa}\kappa 
v_{S_{1}}v_{S_{2}}}{\sqrt{2}v_{S_{3}}}
\end{array}}
\end{equation*}

Diagonalizing the mass-square matrix $\mathcal{M}^2_{{\rm even}}$ yields six mass eigenstates for the CP-even Higgs bosons in the spectrum. Similarly for the CP-odd Higgs fields;

\begin{equation}
\mathcal{M}_{{\rm odd}}^{2} = \left( 
\begin{array}{cccccc}
P_{11}^{2} & P_{12}^{2} & P_{13}^{2} & P_{14}^{2} & P_{15}^{2} & P_{16}^{2} \\
 & P_{22}^{2} & P_{23}^{2} & P_{24}^{2} & P_{25}^{2} & P_{26}^{2} \\
 & & P_{33}^{2} & P_{34}^{2} & P_{35}^{2} & P_{36}^{2} \\
& &  & P_{44}^{2} & P_{45}^{2} & P_{46}^{2} \\
 &  &  &  & P_{55}^{2} & P_{56}^{2} \\
 & &  &  &  & P_{66}^{2} \\
\end{array}
\right)
\end{equation}
and, {the non-zero elements of $\mathcal{M}_{{\rm odd}}^{2}$ are} 
\begin{equation*}
\setstretch{2}
\begin{array}{l}
P_{11}^{2} = \dfrac{A_{\lambda}\lambda v_{S}v_{u}}{\sqrt{2}v_{d}}~,~ P_{12}^{2} = \dfrac{A_{\lambda}\lambda v_{S}}{\sqrt{2}}~,~ P_{13}^{2} = 
\dfrac{A_{\lambda}\lambda v_{u}}{\sqrt{2}}~, \\
P^{2}_{22} = \dfrac{A_{\lambda}\lambda v_{d}v_{S}}{\sqrt{2}v_{u}}~, P^{2}_{23}\dfrac{A_{\lambda}\lambda v_{d}}{\sqrt{2}}~, \\
P^{2}_{33} = \dfrac{1}{2v_{S}}\left(-2m_{SS_{1}}^{2}v_{S_{1}}-2m_{SS_{2}}^{2}v_{S_{2}}+\sqrt{2}A_{\lambda}\lambda v_{d}v_{u}  \right)~,~P^{2}_{34}=-m^{2}_{SS_{1}}~,~P^{2}_{35}=-m^{2}_{SS_{2}}~, \\
P^{2}_{44}=\dfrac{1}{2v_{S_{1}}} \left(-2m_{SS_{1}}^{2}v_{S}+\sqrt{2}A_{\kappa}\kappa v_{S_{2}}v_{S_{3}} \right)~,~ p^{2}_{45}=\dfrac{A_{\kappa}\kappa v_{S_{3}}}{\sqrt{2}}~,~P^{2}_{46}=\dfrac{A_{\kappa}\kappa v_{S_{2}}}{\sqrt{2}} \\
P^{2}_{55} = \dfrac{1}{2v_{S_{2}}}\left(-2m_{SS_{2}}^{2}v_{S} + \sqrt{2}A_{\kappa}\kappa v_{S_{1}}v_{S_{3}}  \right)~,~P^{2}_{56}=\dfrac{A_{\kappa}\kappa v_{S_{1}}}{\sqrt{2}}~,~P^{2}_{66}=\dfrac{A_{\kappa}\kappa v_{S_{1}}v_{S_{2}}}{\sqrt{2}v_{S_{3}}}
\end{array}
\end{equation*}

When the mass-square matrix of the CP-odd Higgs fields are diagonalized, two eigenstates out of six happen to be the massless Goldstone bosons, and thus there remain four CP-odd Higgs bosons in the mass spectrum. 

As can be seen from the mass-square matrices above, the MSSM Higgs fields 
and the MSSM singlet scalars of the secluded $U(1)'$ model non-trivially mix in forming the physical Higgs boson states. Such a mixing can yield non-SM 
Higgs bosons of light mass, which can potentially lead signals in the collider experiments \cite{Un:2016hji}. In this context profiling the Higgs bosons in the spectrum is of importance in  constraining the allowed parameter space of the model. If a Higgs boson mass state, except the SM-like 
state, is formed mostly by the MSSM Higgs fields,  then the current constraints from rare $B-$meson decays such as $B_{s}\rightarrow \mu^{+}\mu^{-}$ and $B\rightarrow X_{s}\gamma$ bound their masses at about 400-500 GeV \cite{Babu:2020ncc,Gomez:2020gav,Raza:2018jnh}. Thus, if the spectrum involves Higgs bosons lighter than the SM-like Higgs boson are excluded by 
these constraints if they are significantly formed by the MSSM Higgs fields. 

Even though the constrained mentioned above can distinguish the MSSM Higgs fields from the MSSM singlet scalars, they can still interfere through their 
non-trivial mixing with the MSSM Higgs fields. First, such mixing can allow some decay modes of the light Higgs bosons into the SM particles, which potentially yield a signal at low mass scales. Besides, since the mixing induce a tree-level coupling with the SM-like Higgs boson, the light Higgs boson states can enhance the invisible decays of the SM-like Higgs bosons. One can avoid such inconsistencies by constraining the decay modes of these light Higgs bosons into the SM particles, and the invisible Higgs decays as ${\rm BR}(h\rightarrow ~ {\rm invisible}) \lesssim 10\%$ \cite{CMS:2019bke,Aaboud:2018sfi,Sirunyan:2018owy,Aad:2015pla,Chatrchyan:2008aa,Khachatryan:2016whc}. 

\section{Scanning Procedure and Experimental Constraints}
\label{sec:scan}

We have employed SPheno 4.0.4 package \cite{Porod:2003um,Porod:2011nf,Goodsell:2014bna} generated with SARAH 4.14.3 \cite{Staub:2013tta,Staub:2015kfa,Goodsell:2014bna}. In this package, the weak scale values of the gauge 
and Yukawa couplings are evolved to the unification scale $M_{{\rm GUT}}$ 
via the renormalization group equations (RGEs). $M_{{\rm GUT}}$ is determined by the requirement of the gauge coupling unification, described as $g_{3}\approx g_{2}=g_{1}=g'_{1}$, where $g_{3}$, $g_{2}$ and $g_{1}$ are the MSSM gauge couplings for $SU(3)_{C}$, $SU(2)_{L}$ and $U(1)_{Y}$ respectively, while $g'_{1}$ corresponds to the gauge coupling for $U(1)'$. Concerning the contributions from the threshold corrections to the gauge couplings at $\mgut$ arising from some unknown breaking mechanisms of the GUT gauge group, $g_{3}$ receives the largest contributions \cite{Hisano:1992jj}, and it is allowed to deviate from the unification point up to about $3\%$. If a solution does not satisfy this condition within this allowance, SPheno does not generate an output for such solutions by default. Hence, the existence of an output file guarantees that the solutions {are} compatible with the unification condition, and $g_{3}$ deviates no more than $3\%$. {After $M_{{\rm GUT}}$ is calculated, all the SSB parameters, determined with the boundary conditions at $M_{{\rm GUT}}$, along with the gauge and Yukawa couplings are evolved back to the weak scale.}

We performed random scans over the parameter space,  shown in Table \ref{tab:scan_lim},  with the universal boundary conditions. Here $m_0$ denotes the spontaneous symmetry breaking (SSB) mass term for all the scalars, while $M_{1/2}$ stands for the SSB mass terms for the gauginos including the one associated with the U(1)$^\prime$ gauge group. $\tan \beta$ is the ratio of VEVs of the MSSM Higgs doublets, and $A_0$ is the SSB trilinear scalar interacting term,  $\lambda$ is the coupling associated with the interaction of $\hat{H}_u$, $\hat{H}_d$ and $\hat{S}$ fields while $\kappa$ is the coupling of the interaction of $\hat{S}_1$, $\hat{S}_2$ and $\hat{S}_3$ fields. Trilinear couplings for $\lambda$ and $\kappa$ are defined as $\lambda A_\lambda$ and  $\kappa A_\kappa$, respectively at the GUT scale. $h_\nu$ is the Yukawa coupling of the term $\hat{L} \hat{H}_u \hat{N}$.

\begin{table}[h]
	\setlength\tabcolsep{20pt}
	\renewcommand{\arraystretch}{2.0}
	\begin{tabular}{|c|c||c|c|}
		\hline
		Parameter      & Scanned range& Parameter      & Scanned range \\
		\hline
		$m_0$        & $[0, 10]$ TeV     & $v_S$   & $[1, 20]$ TeV  \\
		$M_{1/2}$   & $[0, 10]$ TeV     & $v_{S_1}$   & $[3, 20]$ TeV  \\
		$\tan\beta$   & $[1, 50]$        & $v_{S_2}$  & $[3, 20$] TeV  \\
		$A_0/m0$   & $[-3, 3]$   & $v_{S_3}$  & $[3, 20]$ TeV \\		
		$\lambda$   & $[0.01, 0.5]$ & $A_\lambda$   & $[0, 10]$ TeV  \\		
		$\kappa$   & $[0.1, 1.5]$ & $A_\kappa$   & $[-10, 0]$ TeV   \\	
		$h_\nu $   & $[10^{-11}, 10^{-7}]$  & -   & - \\																						
		\hline
	\end{tabular}
	$\vspace{0.5cm}$
	\caption{\label{tab:scan_lim} Scanned parameter space.}
\end{table}

In analysing the data and implications of the model, we impose the LEP2 bounds on the charged particles such that the model does not yield any new 
charged particles whose mass is lighter than about 100 GeV \cite{Patrignani:2016xqp}. In addition, since it has been significantly being updated, we require the consistent solutions yield gluino mass as $m_{\tilde{g}} \ge 2100$ GeV. Another important mass bound comes from the Higgs boson. We 
require one of the Higgs bosons in solutions to exhibit the SM-like Higgs boson properties in terms of its mass and decay channels {reported by the ATLAS \cite{ATLAS:2012yve,ATLAS:2017bxr,ATLAS:2019aqa,ATLAS:2018uoi} and CMS \cite{CMS:2012qbp,CMS:2017dib,CMS:2015hra,CMS:2013zmy} collaborations}. Including the scalars, whose VEVs break the $U(1)'$ symmetry, the low scale spectrum involves six CP-even Higgs boson mass. Since the mixing between the $U(1)'$ breaking scalar fields and the MSSM Higgs fields is expected to be small, 
the SM-like Higgs boson should be formed mostly by the MSSM Higgs fields. 
In this context, the SM-like Higgs boson needs be identified not only with its mass, but also its mixing. If a solution yield one of the Higgs bosons ($h_{i},~i=1,\ldots 6$) with a mass of about 125 GeV \cite{Khachatryan:2016vau}, we also require $|ZH(i,1)|^{2} +|ZH(i,2)|^{2} \gtrsim 80\%$, {where $Z_{H}$ matrix quantifies the mixing among the Higgs bosons}. 

{Another one of the important constraints arises from the REWSB conditions \cite{Ibanez:1982fr,Inoue:1982pi,Ibanez:1982ee,Ellis:1982wr,AlvarezGaume:1983gj} which requires the $\mu-$term consistent with EWSB}.
 We also implement the constraints from rare $B$-meson decays such as $ {\rm BR}(B \rightarrow X_{s} \gamma) $ \cite{Amhis:2012bh}, $ {\rm BR}(B_s \rightarrow \mu^+ \mu^-) $ \cite{Aaij:2012nna} and $ {\rm BR}(B_u\rightarrow\tau \nu_{\tau}) $ \cite{Asner:2010qj}. Then, we require that the predicted relic density of the neutralino LSP agrees within 5$ \sigma $ with the  recent Planck  results \cite{Aghanim:2018eyx}. The relic density of the LSP and scattering cross sections for direct detection experiments are calculated with $\mo$ (version 5.0.9) \cite{Belanger:2018mqt}. The experimental constraints are summarized in Table \ref{tab:constraints}. The following  list  summarizes the relation between colours and constraints imposed in our forthcoming plots.

\begin{itemize}
	\item Grey: {Represents the points compatible with the Radiative EWSB (REWSB) and  neutralino LSP},
	
	\item Blue: {Forms a subset of grey and represents points satisfying the constraints on the SUSY particle masses, Higgs boson mass and its couplings, and $B-$physics constraints,}
	
	\item Red: {Forms a subset of blue and represents the points which are consistent with the Planck bounds on the relic density of LSP neutralino within $5\sigma$ together with other constraints mentioned for blue points.}
	
\end{itemize}

\begin{table}
\caption{The experimental constraints employed in our analyses.}
\centering
\setstretch{2.0}
\begin{tabular}{|c|c|c|}
\hline
Observable & Constraint & Ref. \\ \hline
$m_{h}$ & $[122-128]$ GeV & \cite{ATLAS:2012yve,CMS:2013btf} \\
$M_{Z^{\prime}}$ & $\geq 4$ TeV & \cite{Pankov:2019yzr,Bobovnikov:2018fwt,CMS:2016wev,Lao:2020inc,ATLAS:2019erb,CMS:2019tbu} \\
$m_{\tilde{g}}$ & $\geq 2.1$ TeV & \cite{ATLAS:2019fag} \\
$m_{\tilde{\chi}_{1}^{\pm}},m_{\tilde{\tau}}$ & $\geq 100$ GeV & \cite{Patrignani:2016xqp} \\
${\rm BR}(B\rightarrow X_{s}\gamma)$ & $[2.99 - 3.87]\times 10^{-4}~(2\sigma)$ & \cite{Amhis:2012bh} \\
${\rm BR}(B_{s}\rightarrow \mu^{+}\mu^{-})$ & $[0.8 - 6.2]\times 10^{-9}~(2\sigma)$ & \cite{Aaij:2012nna} \\
$\dfrac{{\rm BR}(B_{u}\rightarrow \tau \nu_{\tau})_{{\rm Secluded~}U(1)^{\prime}}}{{\rm BR}(B_{u}\rightarrow \tau \nu_{\tau})_{{\rm SM}}}$ & $[0.15 - 2.41]~(2\sigma)$ & \cite{Asner:2010qj} \\
$\Omega_{{\rm CDM}}h^{2}$ & $[0.114 - 0.126]~(5\sigma)$ & \cite{Aghanim:2018eyx}\\ \hline
\end{tabular}
\label{tab:constraints}
\end{table}

\section{Results}
\label{sec:results}

\begin{figure}[t!]
	\centering
	\includegraphics[scale=0.4]{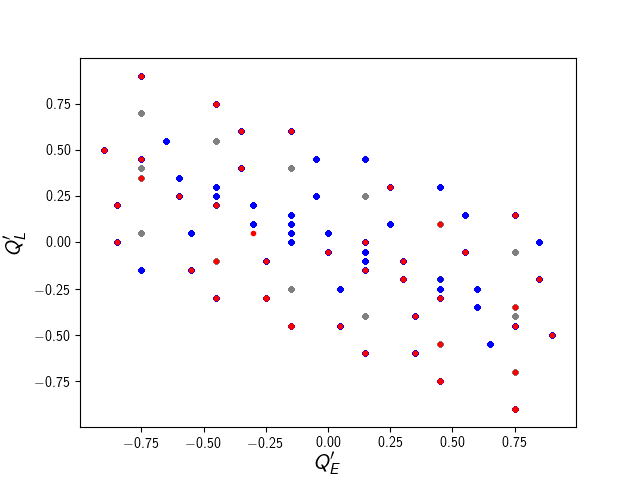} 
	\includegraphics[scale=0.4]{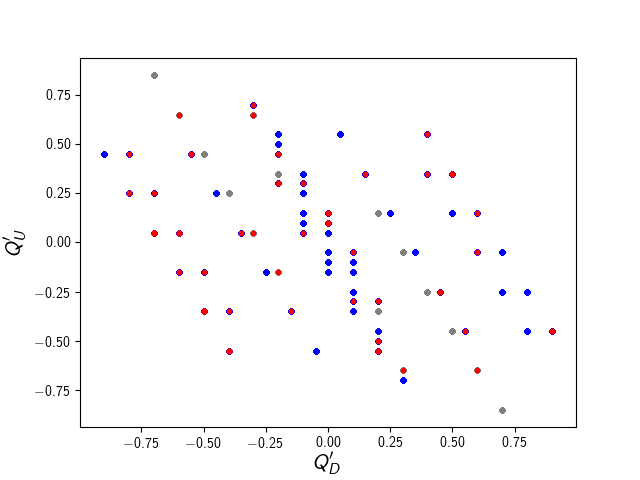} \\	
	\includegraphics[scale=0.4]{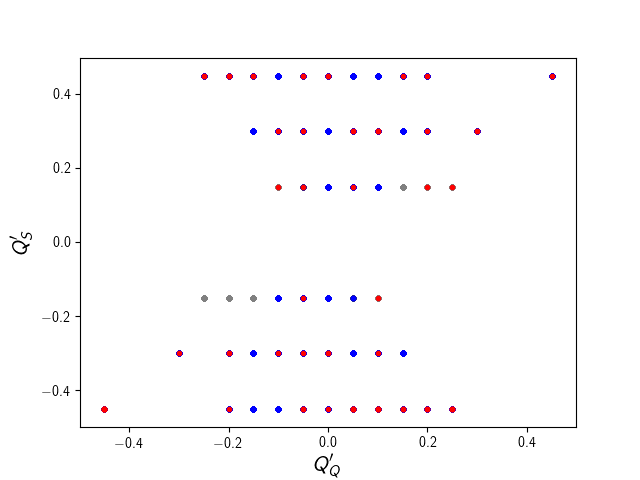}  \\	
	\caption{The distributions of the U(1)$^\prime$ charges in secluded U(1)$^\prime$ model allowed by various theoretical and experimental conditions from Section \ref{sec:scan} over the following planes: $ Q'_{L}-Q'_{E}$, $ Q'_{U} -Q'_{D}$ and $ Q'_{S} -Q'_{Q}$. The colour convention is as listed at the end of Section \ref{sec:scan}. }
	\label{fig:Charges}
\end{figure}

In this section we present our results in light of the constraints {discussed} in the previous section. First, we focus on the U(1)$^\prime$ charges which characterizes the secluded U(1)$^\prime$  model. Fig. \ref{fig:Charges} depicts the  U(1)$^\prime$ charge sets satisfying various theoretical and experimental bounds. The colour convention is as listed at the end of Section \ref{sec:scan}. Herein, we show  charges  for left and right chiral fermions and MSSM singlet by visualising our scan points over the planes $ Q'_{L}-Q'_{E}$ and $ Q'_{U} -Q'_{D}$ (top panels), and $ Q'_{S}-Q'_{Q}$ (bottom panel). {Note that the charges are normalized to unity. As can be seen from the top left panel, the constraints allow a large number of different solution sets, and a wide range for the charges can be accommodated, e.g., the $Q^{\prime}_{L},Q^{\prime}_{E} \lesssim |0.9|$}. As for quarks, the right handed up-type  quark  charges ($Q^{\prime}_{U}$) and the right handed down-type ones ($Q^{\prime}_{D})$
{exhibits} almost the same behaviour (as shown in the top right panel of the figure). Furthermore, it can easily be read, from the bottom panel, {that} $Q^{\prime}_{S}$ charge is always far away from zero since $Q^{\prime}_{S}=-(Q^{\prime}_{H_u}+Q^{\prime}_{H_d})$. After applying all theoretical conditions and experimental constraints, $Q^{\prime}_{S}$ an $Q^{\prime}_{Q}$ charges are restricted to certain regions, $Q^{\prime}_{S},Q^{\prime}_{Q} \lesssim |0.5|$. Since there is not  a direct anomaly cancellation condition between $Q^{\prime}_{Q}$ and $Q^{\prime}_{S}$, it is possible to find various $Q^{\prime}_{Q}$ values for the fixed values of $Q^{\prime}_{S}$.  

\begin{figure}[t!]
	\centering
	\includegraphics[scale=0.4]{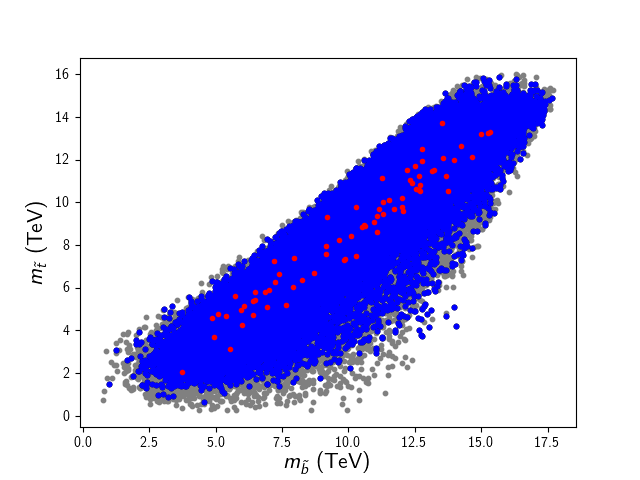} 
	\includegraphics[scale=0.4]{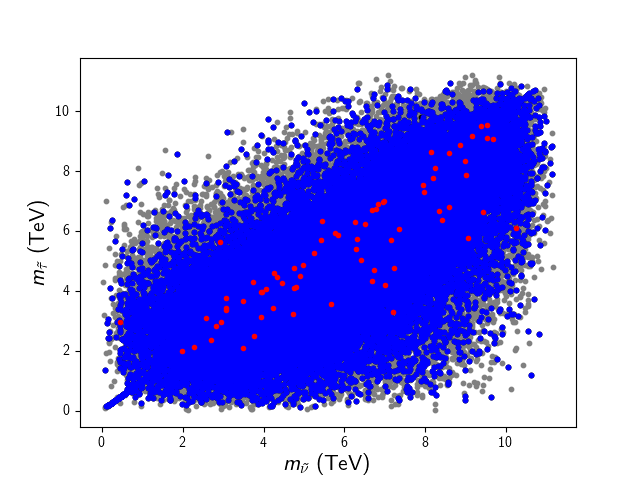} \\	
	\caption{The mass spectrum of SUSY particles over the following planes:  $m_{\tilde{t}}-m_{\tilde{b}}$ (left) and $m_{\tilde{\tau}}-m_{\tilde{\nu}}$ (right). The colour convention is as listed at the end of Section \ref{sec:scan}.}
	\label{fig:spar}
\end{figure}

Fig. \ref{fig:spar} displays the the mass spectrum of SUSY particles in $m_{\tilde{t}}-m_{\tilde{b}}$ (left) and $m_{\tilde{\tau}}-m_{\tilde{\nu}}$ (right) planes. The colour convention is as listed at the end of Section \ref{sec:scan}.  The left panel shows that sbottom and  stop masses are heavy in general and should be $ 3 \ {\rm TeV} \lesssim m_{\tilde{b}},m_{\tilde{t}} \lesssim 15 \ {\rm TeV}$. {Even though these mass scales are far beyond the reach of the current LHC experiments}, they can be probed in the future collider searches \cite{Cohen:2013xda,Altin:2020qmu}. {Similarly, the right panel also reveals that stau can be as light as only 2 TeV, compatible with all experimental bounds. Even though the sneutrino mass can be realized as low as about 500 GeV, it is, in general, heavier than stau for most of the solutions compatible with all experimental bounds.}

\begin{figure}[h]
	\centering
	\includegraphics[scale=0.45]{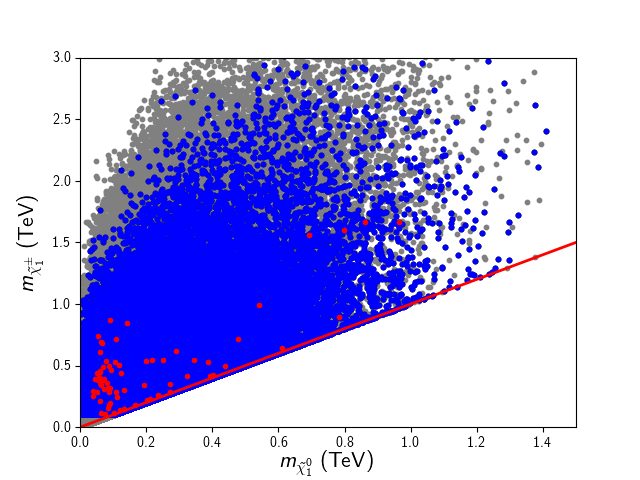}
	\includegraphics[scale=0.45]{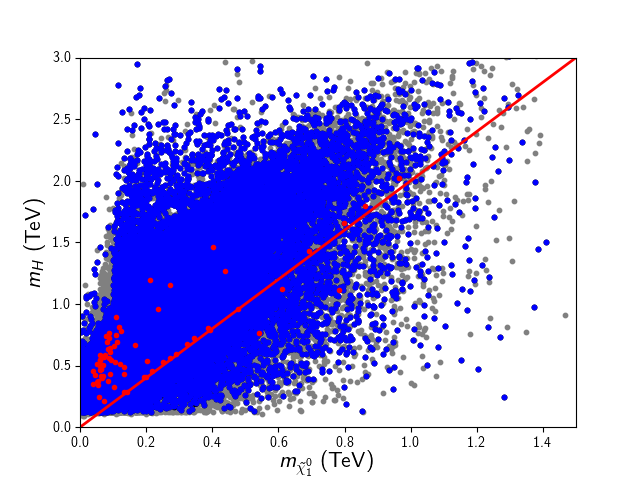}
	\includegraphics[scale=0.45]{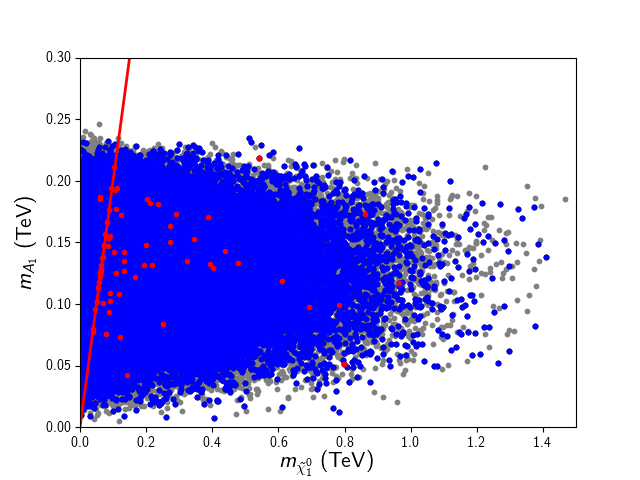} 
	\includegraphics[scale=0.45]{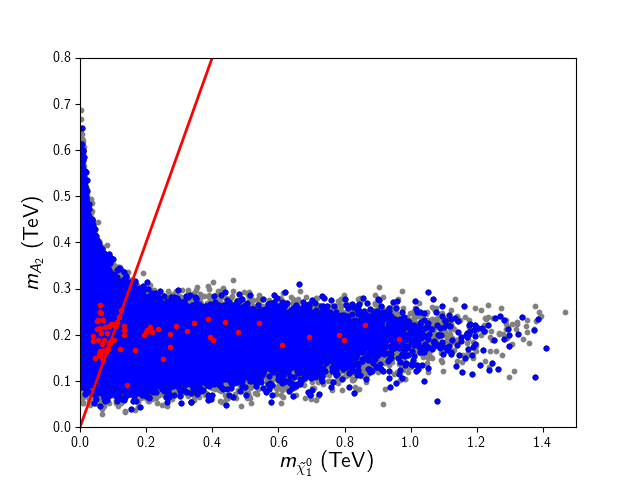}
	\caption{The mass spectrum of the lightest neutralino and chargino and relic density channels over the following planes:  $m_{\tilde{\chi}_{1}^{\pm}}-m_{\tilde{\chi}_{1}^{0}}$ (top left), 
		$m_{H}-m_{\tilde{\chi}_{1}^{0}}$ (top right), 
		$m_{A_{1}}-m_{\tilde{\chi}_{1}^{0}}$ (bottom left) and $m_{A_{2}}-m_{\tilde{\chi}_{1}^{0}}$ (bottom right). The colour convention is as listed at the end of Section \ref{sec:scan}.}
	\label{fig:neutralino}
\end{figure}

Fig. \ref{fig:neutralino} shows the neutralino and chargino mass spectrum {with diagonal lines emphasizing the co-annihilation and annihilation channels of LSP neutralino} in $m_{\tilde{\chi}_{1}^{\pm}}-m_{\tilde{\chi}_{1}^{0}}$ (top left), 
$m_{H}-m_{\tilde{\chi}_{1}^{0}}$ (top right), 
$m_{A_{1}}-m_{\tilde{\chi}_{1}^{0}}$ (bottom left) and $m_{A_{2}}-m_{\tilde{\chi}_{1}^{0}}$ (bottom right) planes. The colour coding is the same as in Fig. \ref{fig:spar}. {As is shown in the $m_{\tilde{\chi}_{1}^{\pm}}-m_{\tilde{\chi}_{1}^{0}}$ plane, the chargino and neutralino can be as light as about 50-100 GeV. Even though the LSP neutralino mass can be realized, in principle, lighter than 50 GeV, the mass scales below 50 GeV trigger the invisible decays of the SM-like Higgs boson; thus, we consider the solutions with $m_{\tilde{\chi}_{1}^{0}} \lesssim 50$ GeV to be excluded. Similarly, the chargino masses lower than 103.5 GeV are excluded as required by the LEP results. Apart from the lower bounds, the LSP neutralino happens mostly to be lighter than about 1 TeV. As discussed before, the MSSM neutralinos cannot be consistent if their masses are lighter than about 500 GeV due to the severe constraints from rare $B-$meson decays and their relic density. Thus the solutions with $m_{\tilde{\chi}^{0}_{1}} \lesssim 500$ GeV should lead to LSPs which are formed mostly by the MSSM singlet fields. The chargino can be as heavy as about 1.5 TeV in the consistent spectra, while its mass can also be at the order of $\mathcal{O}(100)$ GeV. The light chargino solutions are expected to be formed mostly by Higgsinos because of a sub-TeV scale $\mu-$term. In this case, if the LSP is formed by Singlinos, while the lightest chargino is mostly a Higgsino, then one can identify the chargino-neutralino coannihilation scenario, through the interactions among the MSSM Higgsinos and $U(1)'$ Singlinos, in the approximate mass-degeneracy region represented with the diagonal line in the $m_{\tilde{\chi}_{1}^{\pm}}-m_{\tilde{\chi}_{1}^{0}}$ plane. In this region, the LSP neutralino coannihilate together with the lightest chargino which leads to lower the relic density of the LSP. Since the solutions compatible with the Planck bound are mostly accumulated around the diagonal line, the chargino-neutralino coannihilation scenario is required by the consistent DM solutions when $m_{\tilde{\chi}_{1}^{\pm}}\simeq m_{\tilde{\chi}_{1}^{0}} \lesssim 0.75$ TeV.}

Even though one can realize consistent DM solutions through chargino-neutralino coannihilation scenario, the $m_{\tilde{\chi}_{1}^{\pm}}-m_{\tilde{\chi}_{1}^{0}}$ plane presents other solutions out of the mass degeneracy region (red points far above the diagonal line). These solutions cannot be identified in the chargino-neutralino coannihilation scenario; thus, the relic density of LSP neutralino should be lowered in other coannihilation and/or annihilation scenarios. Since other SUSY particles are either heavy (as stop, sbottom and stau shown in Fig. \ref{fig:spar}) or they do not directly couple to the Singlino LSP (as sneutrino), the relic density of LSP neutralino is most likely lowered by its annihilation processes into a neutral Higgs boson as displayed in the $m_{H}-m_{\tilde{\chi}_{1}^{0}}$, $m_{A_{1}}-m_{\tilde{\chi}_{1}^{0}}$ and $m_{A_{2}}-m_{\tilde{\chi}_{1}^{0}}$ planes of Fig. \ref{fig:neutralino}. The diagonal lines in these planes indicates the regions where $2m_{\tilde{\chi}_{1}^{0}}=m_{H},m_{A_{1}},m_{A_{2}}$ respectively. In the regions represented by the diagonal lines, $m_{H}$ can be as light as about 100 GeV, while it can also be realized as heavy as about 2 TeV. On the other hand, the lighter CP-odd Higgs boson masses are found to be bounded as $m_{A_{1}}\lesssim 300$ GeV and $m_{A_{2}}\lesssim 600$ GeV, as shown in the bottom planes of Fig. \ref{fig:neutralino}. One can conclude from such results that the LSP neutralino annihilations through the Higgs portal plays an important role to identify consistent DM solutions. Especially the annihilation processes involving CP-odd Higgs bosons significantly lower the relic density of LSP. Recall that the MSSM Higgs bosons contribute to rare $B-$meson decays at these mass scales and violate the constraints from $B-$physics. Thus, these light Higgs bosons should be formed mostly by the MSSM singlet scalars to be consistent with the constraints from rare $B-$meson decays. Besides, a MSSM singlet Higgs boson can strongly couple to the LSP neutralino and significantly lower its relic density. {Even tough these light CP-odd Higgs bosons have MSSM singlet nature, they can interfere in the SM-like Higgs boson decays through $h_{1}\rightarrow A_{i}A_{i}$, if they are formed by the MSSM singlet field $S$, which is allowed to interact with the MSSM Higgs fields at tree level through the coupling $\lambda$. If $\lambda$ is considerably large, it also enhance the mixing between the MSSM Higgs fields and the MSSM singlet $S$ field. However, such a large mixing receives a strong impact from rare $B-$meson decays, especially from $B_{s}\rightarrow \mu\mu$. We observed that these light CP-odd Higgs bosons can lead to ${\rm BR}(h_{1}\rightarrow A_{1}A_{1}) \lesssim 15\%$. However, such solutions are excluded by the Planck bound on the relic abundance of LSP neutralino, which favors  ${\rm BR}(h_{1}\rightarrow A_{1}A_{1}) \sim 0$. In this context, these light CP-odd Higgs bosons can escape from the current analyses which considers the rare and exotic decays of the SM-like Higgs boson \cite{CMS:2019idx,CMS:2018lce,CMS:2017nmj,CMS:2018nsh,ATLAS:2021edm,Curtin:2013fra,Bernon:2014nxa,Cici:2019zir}.}

\begin{figure}[h!]
\centering
\subfigure{\includegraphics[scale=0.45]{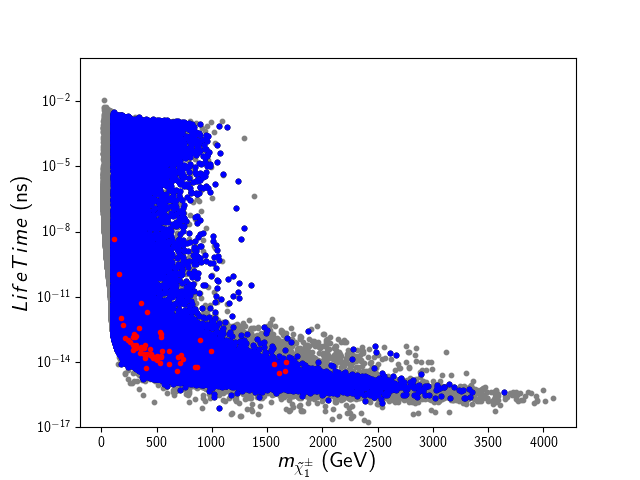}}%
\subfigure{\includegraphics[scale=0.45]{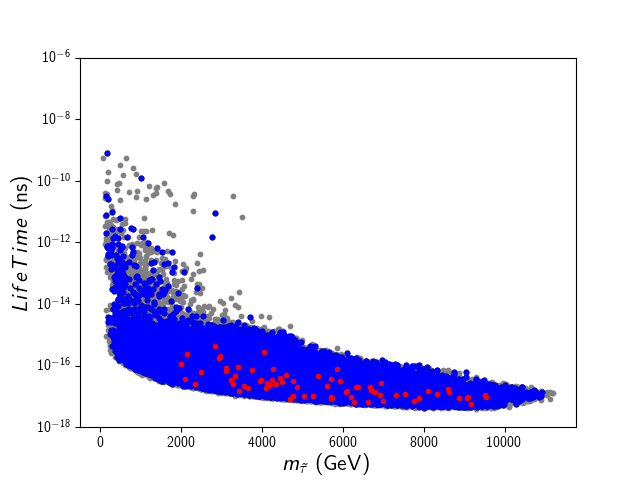}}
\caption{Life-time of charged particles such as the chargino (left) and stau (right) in correlation with their masses. The colour convention is as listed at the end of Section \ref{sec:scan}.}
\label{fig:LT}
\end{figure}

{If the spectrum involves charged particles which are nearly degenerate with the LSP neutralino in mass, the current analyses \cite{ATLAS:2021ttq,ATLAS:2020wjh} yield a strong impact on their life-time. In the models under phenomenological concern, the stability of such light charged particles can be controlled by the tree-level coupling between NLSP and LSP. Even though the current experimental constraints prevent the strongly interacting charged particles to be NLSP, the chargino and the sleptons are, in principal, allowed to be NLSP. As discussed before, since the universal SSB gaugino mass terms do not allow the Wino to be lighter than Bino, the light chargino solutions can be observed only if they are formed by the MSSM Higgsinos, which couple to the MSSM singlet LSP neutralino at tree-level. In addition, the universal SSB mass term for the scalar supersymmetric particles lead to stau to be the lightest slepton in the low scale mass spectrum. However, there is no tree-level coupling between the stau and MSSM singlet LSP neutralino, and the NLSP stau solutions more likely imply stable staus. Such solutions are strictly excluded by the collider analyses, since they would signal in the collisions as a missing charge. We have plotted the life-time of these charged particles in Figure \ref{fig:LT} in correlation with their masses. All points are compatible with REWSB and the LSP neutralino condition. The blue points are consistent with the mass bounds and constraints from rare $B-$meson decays. The red points form a subset of blue and they satisfy the Planck bound on the relic abundance of LSP neutralino within $5\sigma$. The left panel shows the life-time of lighter chargino, which is Higgsino-like. All solutions yield chargino life-time shorter than about $10^{-2}$ ns and compatible with its current bound \cite{ATLAS:2021ttq}. Furthermore, the solutions compatible with the Planck bound on the relic abundance of LSP neutralino (shown in red) can have the charginos, whose life-time cannot be longer than about $10^{-8}$ ns. Similarly the right panel display the results for the stau life-time, which undergoes immediate decay in the solutions within our data, since its life-time is realized to be always shorter than about $10^{-9}$ ns. This is because the stau is always heavier than some other MSSM particles (such as Bino and Higgsinos), which couple to stau at tree-level. The DM constraint shorten its life-time further as $\tau_{\tilde{\tau}} \lesssim 10^{-15}$ ns as seen from the red points. Note that our plots are insensitive to the mass difference of about $1-2$ GeV due to the point size used in plotting. Even though the $m_{\tilde{\chi}_{1}^{\pm}}-m_{\tilde{\chi}_{1}^{0}}$ in Figure 3 has some red points on the diagonal line, these solutions still lead to $1-2$ GeV mass difference between the chargino and LSP neutralino, and such charginos undergo the $\tilde{\chi}_{1}^{\pm}\rightarrow q\bar{q}^{\prime}\tilde{\chi}_{1}^{0}$ decay processes after they are being produced at the colliders, where $q$ and $q^{\prime}$ denote different quarks as required by the electric charge conservation.}

{In addition to discussions about the coannihilation channels in Figure \ref{fig:neutralino}, each species of neutralinos yield different phenomenology and implications in the dark matter experiments.} If the LSP mass eigenstate $\tilde\chi_1^0$ is given in terms of interaction eigenstates by the following linear combination {using the same basis as the neutralino mass matrix given in Eq.(\ref{eq:Mneutralino});}

\begin{equation}
\displaystyle\tilde\chi_1^0=Z_{11}\widetilde{B}+Z_{12}\widetilde{W}^3+Z_{13}\widetilde{H}_d^0+Z_{14}\widetilde{H}_u^0+Z_{15}\widetilde{S}+Z_{16}\widetilde{B}'+Z_{17}\widetilde{S}_1+Z_{18}\widetilde{S}_2+Z_{19}\widetilde{S}_3
\label{eqn:neu_comp}
\end{equation} 
{where $Z_{ij}$ are elements of the diagonalization matrix encoding the possible mixtures in comprising the neutralino mass eigenstates, $\sum |Z_{ij}|^2=1$ by the normalization condition and $|Z_{1j}|^{2}$ measures the fraction of the $j^{{\rm th}}$ particle in the composition of LSP neutralino.} 


{The linear superposition of LSP neutralino given in Eq.(\ref{eqn:neu_comp}) implies that $U(1)^{\prime}$ can deviate the LSP neutralino from the MSSM phenomenology by the total fraction of $U(1)^{\prime}$ particles expressed as $|Z_{15}|^{2}+|Z_{16}|^{2}+|Z_{17}|^{2}+|Z_{18}|^{2}+|Z_{19}|^{2}$. If  $|Z_{17}|^{2}+|Z_{18}|^{2}+|Z_{19}|^{2}$ dominates the other elements of LSP mass diagonalization matrix, then the dark matter is realized to be mostly decouple from the other particles. In this case, the current sensitivity of the experiments in direct and indirect searches of dark matter cannot probe such solutions. On the other hand, if $|Z_{15}|^{2}$ is significantly larger, then the dark matter is again composed mostly by the MSSM singlet, i.e. $\tilde{S}$, which interacts with the MSSM particles through the Higgs portal. The current and projected sensitivity of the direct detection of dark matter experiments can provide a potential probe for such solutions. Moreover, even if it does not form the dark matter significantly, it can still alter the dark matter implications through its mixing with the MSSM neutralinos. Finally, even though $\tilde{B}^{\prime}$ is also theoretically allowed to form the dark matter, the mass is mostly controlled by $v_{S}$. Since the heavy mass bound on $Z^{\prime}$ bounds $v_{S}$ at about a few TeV from below, the neutralino mass eigenstate, which is mostly formed by $\tilde{B}^{\prime}$, is found rather to be heavy.}

\begin{figure}[h]
	\centering
	\includegraphics[scale=0.45]{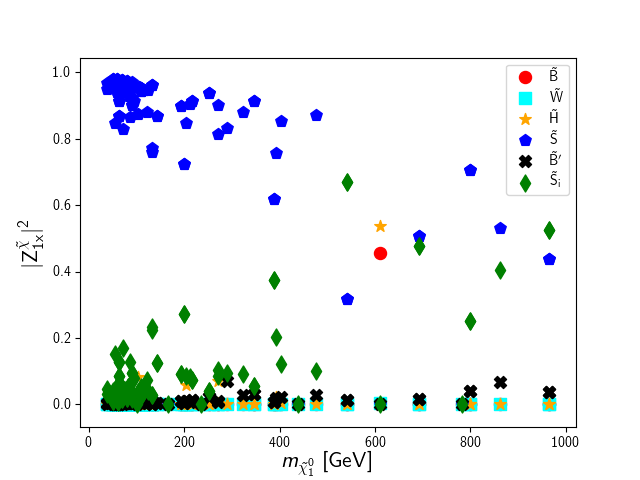} 
	\includegraphics[scale=0.45]{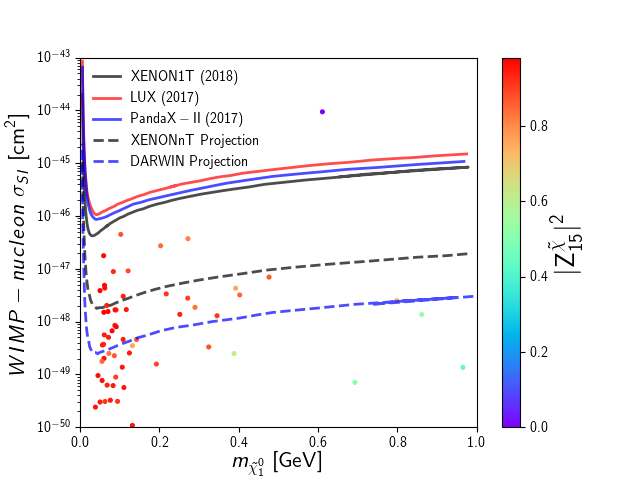} \\	
	\caption{The composition of the LSP versus its mass (left) and DM-nucleon SI scattering cross section as a function of the  mass of the lightest neutralino LSP (right). Limits from current (solid) and future (dashed) experiments are also shown.}
	\label{fig:DarkMatter1}
\end{figure}

{We can summarize the discussion about the LSP neutralino composition and its testable implications in the dark matter experiments with plots given in Fig.\ref{fig:DarkMatter1}. {In the left panel, we visualize the branching fraction of each neutralino by using different colors and shapes with the mass of the LSP neutralino. The color and shape convention is given in the legend.} The points represented in this plane are selected such that they are allowed by all the constraints including the Planck bound on the LSP neutralino relic density. As is seen from the blue {pentagons}, the light LSP masses ($\lesssim 350$ GeV) can be realized when the LSP neutralino is formed mostly by $\tilde{S}$ ($\gtrsim 80\%$). $\tilde{S}$ still plays a crucial role for relatively {\color{black}large} mass scales, since its mixing in the LSP composition is realized at about $40\%$ and more for $m_{\tilde{\chi}_{1}^{0}} \gtrsim 400$ GeV. Most of these solutions also reveal that the rest of the LSP neutralino is formed by the other MSSM singlets i.e. $\tilde{S}_{1}$, $\tilde{S}_{2}$ and $\tilde{S}_{3}$. In this context, the dark matter is realized to be almost a MSSM singlet, while it can interact with the MSSM particles through the Higgs portal. As discussed before, the fraction for $\tilde{B}^{\prime}$ is realized as large as only about $10\%$ and less (black {crosses} in the left panel of Fig. \ref{fig:DarkMatter1}).} 

{The MSSM neutralinos become effective in the LSP composition when $m_{\tilde{\chi}_{1}^{0}}\gtrsim 600$ GeV, which can be measured as about $50\%$ Bino fraction (red 1), and about $25\%$ for each MSSM Higgsino (turquois 3 and orange 4). This mass scale bounding the Bino-Higgsino mixture is a direct result of the gluino mass bound and the Planck bound on the relic density of LSP neutralino. Since we employ universal gaugino masses at the GUT scale, the gluino mass bound excludes the region where $M_{1/2} \lesssim 600$ GeV. In addition, when the Higgsinos form the LSP neutralino, the relic density constraint is satisfied when $m_{\tilde{\chi}_{1}^{0}}\gtrsim 700$ GeV (see, for instance, \cite{Raza:2018jnh,Baer:2012by,Delgado:2020url}). The Higgsino fraction is also constrained by the results from the direct detection experiments, since it yields large cross-sections for the dark matter scattering at nuclei.
}

{As is seen from the discussion above, the secluded $U(1)^{\prime}$ model yield solutions in which the dark matter is mostly formed by the Singlino ($\tilde{S}$). Even though there are not many channels in scattering of dark matter at nuclei, these solutions can be traced down in the direct detection experiments through the Higgs portal, and their signature can be significant depending on how strongly it interacts with the MSSM Higgs fields. The right panel of Fig.\ref{fig:DarkMatter1} show the results for the spin independent scattering cross-section of dark matter with respect to its mass. The represented solutions are selected to be consistent with the constraints employed in our analyses. The experimental results from the direct detection experiments are represented with the curves. The black, blue and red solid lines show XENON1T \cite{XENON:2018voc}, PandaX-II \cite{PandaX-II:2017hlx} and LUX \cite{LUX:2016ggv} upper limits for the  SI  ${\tilde\chi}_1^0 $ - nucleon cross section, respectively, while the black and blue dashed lines illustrate the prospects of the XENONnT and DARWIN for future experiments \cite{DARWIN:2016hyl}, respectively. We also display a color bar which relates the color coding to the Singlino fraction in the composition of LSP. The red points correspond to Singlino-like dark matter in which the Singlino fraction is realized greater than about $70\%$. The solutions with $m_{\tilde{\chi}_{1}^{0}} \lesssim 350$ GeV yielding scattering cross-sections larger than about $2\times 10^{-48}~{\rm cm}^{2}$ can be tested in XENON experiment soon, while those with $\sigma_{{\rm SI}} \in [3\times 10^{49}-2\times10^{48}]~{\rm cm}^{2}$ are expected to be probed by the DARWIN collaboration. DARWIN will also be able to test the Singlino dark matter when the LSP mass is relatively heavier ($\sim 500$ GeV). Finally, we also display a solution exemplifying the  MSSM-like dark matter (blue). These solutions were identified as Bino-Higgsino mixture in the previous discussion, and as is seen from the right plane, the direct detection experiments yield a strong negative impact on such solutions, since it predicts a large scattering cross-section.}

\begin{table}[h!]
\caption{The benchmark points for different scenarios. The points are selected to be consistent with the experimental constraints. All masses are given in TeV, and the cross-sections in cm$ ^{2} $.}
\centering
\setstretch{2.0}
\scalebox{0.73}{
\begin{tabular}{|c|cc|cc|c|}
\hline
& \multicolumn{2}{c|}{Scenario I} & \multicolumn{2}{c|}{Scenario II} & Scenario III \\ \hline
Parameters &  Point1 & Point 2 & Point 3 & Point 4 & Point 5 \\ \hline
$g_1^{\prime}$ &0.36&0.57&0.41&0.41&0.45\\ 
$\tan\beta$ &15.7&21.3&12.3&45.7&20.0\\
			$\mu_{\rm eff}$ &0.110&0.105&0.389&0.419&0.630\\
			$( \lambda,\ \kappa)$&(0.02,\ 0.63)&(0.017,\ 0.72)&(0.09,\ 0.67)&(0.09,\ 1.07)&(0.05,\ 0.51)\\
			$(A_{\lambda},\ A_{\kappa})$ &(4.6,\ -4.2)&(2.2,\ -3.3)&(3.9,\ -3.2)&(5.0,\ -8.6)&(4.8,\ -3.1)\\
			$(v_{s_1},\ v_{s_2},\ v_{s_3}) $ &(15.5,\ 14.6,\ 12.6)&(13.0,\ 8.55,\ 11.8)&(14.6,\ 11.2,\ 14.1)&(11.5,\ 19.0,\ 9.65)&(10.0,\ 10.0,\ 6.8)\\
			$(m_{0})$ &3.835&4.569&9.610&7.267&5.512\\
			$(M_{1/2})$ &3.345&9.359&5.500&3.502&1.450\\
			$(A_{0})$ &1.305&-2.572&1.154&-3.639&-0.305\\ \hline
			$m_{Z^\prime}$ &5.53 &5.19&6.43&5.69&5.62\\
			$(m_{H^0_1},\ m_{H^0_2})$ & 
			(0.1243,\,0.659) &
			(0.127,\,0.325) &
			(0.1228,\,0.242) &
			(0.125,\,0.783)&
			(0.1234,\,1.121)  \\
			$(m_{A^0_1},\ m_{A^0_2})$ & 
			(0.142,\,0.188) &
            (0.210,\,0.223) &
            (0.117,\,0.157) &
            (0.172,\,0.253)& 
            (0.119,\,0.179)\\
            $(m_{\tilde\chi^0_1},\ m_{\tilde\chi^0_2})$ & 
			(0.103,\,0.115) &
			(0.102,\,0.112) &
			(0.056,\,0.408) &
			(0.124,\,0.436)& 
			(0.611,\,0.637)
			\\
			$(m_{\tilde\chi^\pm_1},\ m_{\tilde\chi^\pm_2})$ & (0.117,\,2.783) &(0.113,\,7.813)&(0.409,\,4.638)&(0.438,\,2.973)& (643,\,1.239)
			\\
			$m_{H^{\pm}}$ &2.872&2.201&4.327&9.796 &7.847
			\\
			$(m_{\tilde t_L},\,m_{\tilde c_L},\,m_{\tilde u_L})$ &(5.37,\,6.89,\,7.13)  &(12.1,\,15.4,\,16.3) &(9.59,\,12.8,\,13.7)&(6.61,\,9.22,\,9.53)&(4.58,\,5.92,\,6.38)
            \\
			$(m_{\tilde t_R},\,m_{\tilde c_R},\,m_{\tilde u_R})$ &(6.41,\,6.89,\,7.13) &(14.6,\,15.4,\,16.3) &(12.2,\,12.8,\,13.7) &(7.38,\,9.22,\,9.53) &(4.89,\,5.92,\,6.38)
             \\
			$(m_{\tilde b_L},\,m_{\tilde s_L},\,m_{\tilde d_L})$ &(6.41,\,6.82,\,7.13)  &(14.6,\,15.4,\,16.3) &(12.0,\,12.1,\,13.7)&(7.37,\,9.39,\,9.53)&(4.88,\,5.92,\,6.51)
            \\
            $(m_{\tilde b_R},\,m_{\tilde s_R},\,m_{\tilde d_R})$ &(6.75,\,6.82,\,7.13)  &(15.1,\,15.4,\,16.3) &(12.2,\,12.1,\,13.7)&(7.52,\,9.39,\,9.53) &(6.29,\,5.92,\,6.51)
            \\
			$(m_{\tilde \tau_L},\,m_{\tilde \mu_L},\,m_{\tilde e_L})$ &(3.96,\,4.03,\,4.34)  &(4.27,\,4.58,\,7.80) &(8.87,\,8.91,\,10.2)&(5.03,\,7.55,\,7.57)&(4.87,\,5.16,\,6.24)
			\\
			$(m_{\tilde \tau_R},\,m_{\tilde \mu_R},\,m_{\tilde e_R})$  & (4.31,\,4.03,\,4.34) &(7.71,\,4.58,\,7.80) &(10.1,\,8.91,\,10.2) &(6.41,\,7.55,\,7.57) & (6.12,\,5.16,\,6.24)
			\\
			$(m_{\tilde\nu_{\tau_L}},\, m_{\tilde\nu_{\mu_L}},\,m_{\tilde\nu_{e_L}})$ &(3.97,\,3.97,\,4.34)  &(3.73,\,3.73,\,7.80)&(8.87,\,8.91,\,10.9) &(6.41,\,7.26,\,7.55)&(4.98,\,4.98,\,6.24)
			\\
			$(m_{\tilde\nu_{\tau_R}},\, m_{\tilde\nu_{\mu_R}},\,m_{\tilde\nu_{e_R}})$ &(4.30,\,3.97,\,4.34)  &(7.71,\,3.73,\,7.80) &(10.9,\,8.91,\,10.9)&(7.26,\,7.26,\,7.55)&(6.12,\,4.98,\,6.24)
			\\[0.1em]
			\hline
			$\Omega_{DM}h^2$ &0.1146 &0.1161&0.1253&0.1178&0.1222	\\ 		
			$\sigma_{{\rm SI}}$ &$ 4.59\times 10^{-47}$ &$ 1.69\times 10^{-48}$&$ 3.65\times 10^{-49}$ &$ 2.61\times 10^{-49}$ &$ 9.64\times 10^{-45}$ 	\\ \hline
\end{tabular}}
\label{tab:benchmark1}
\end{table}

{Before concluding, we present six benchmark points in Table \ref{tab:benchmark1} to exemplify our findings. The results discussed above can be classified into three scenarios as grouped in Table \ref{tab:benchmark1}. Scenarios I and II involve the solutions of MSSM singlet LSP, and the MSSM like LSP solutions refer to Scenario III. Scenario I and Scenario II differ from each other in chargino mass. The mass spectra in Scenario I includes charginos nearly mass degenerate with the LSP neutralino, while it is much heavier than the LSP neutralino in Scenario II. The first two points in Table \ref{tab:benchmark1} exemplify the solutions in Scenario I. The correct relic density of LSP neutralino for these solutions is satisfied through the chargino-neutralino coannihilation scenario. Point 1 predicts $\sigma_{{\rm SI}} \simeq 5\times 10^{-46}$ cm$^{2}$ for the spin-independent scattering of DM, which can be tested soon in XENON experiments. Point 2 displays a solution for $A-$resonance. Points 3 and 4 represent the solutions in Scenario II in which the relic density constraint is satisfied only through $A-$resonance. Point 3 yields $\sigma_{{\rm SI}} \simeq 10^{-48}$ cm$^{2}$, and Darwin experiments will be able to test such solutions in near future. Point 4 depicts a solution in which two LSP neutralinos annihilate into the second CP-odd Higgs boson of mass about 250 GeV. Finally Point 5 displays solutions for Scenario III in which the LSP neutralino is formed by the MSSM neutralinos, which is mostly Higgsino in our model. Such solutions lead to very large cross-sections in DM scattering with nuclei ($\sim 10^{-44}$ cm$^{2}$) and they are excluded by the current constraints set by the direct detection experiments. We also list the sets of $U(1)^{\prime}$ charges for these benchmark points in Table \ref{tab:benchcharges}. The solutions in Scenario I favor sets of charges in which the left-handed quark and lepton fields have relatively low charges under the $U(1)^{\prime}$ group. Scenario II depicts a certain magnitudes of the $U(1)^{\prime}$ charges, while their sign could be negative or positive. Scenario III reveals a fact that all the fields are considerably charged under $U(1)^{\prime}$ group if the LSP neutralino is formed by the MSSM neutralinos.}

\begin{table}[h!]
\caption{Sets of $U(1)^{\prime}$ charges for the benchmark points listed in Table \ref{tab:benchmark1}. The charges satisfy  the conditions of gauge invariance (Eq. \ref{eq:gauge_cond})  in the secluded $U(1)^{\prime}$ model described by the superpotential (Eq. \ref{eq:superpot}) and of anomaly cancellation (Eqs. \ref{eq:anomaly_cond} - \ref{eq:anomaly_cond2} ).}
\centering
\setstretch{2.0}
\scalebox{0.73}{
\begin{tabular}{|c|cc|cc|c|}
\hline
& \multicolumn{2}{c|}{Scenario I} & \multicolumn{2}{c|}{Scenario II} & Scenario III \\ \hline
Parameters & Point1 & Point 2 & Point 3 & Point 4 & Point 5 \\ \hline
			$Q_{Q}^{\prime}$ &0.0&-0.05&-0.2&0.2&-0.15\\
			$Q_{U}^{\prime}$ &-0.45&-0.3&0.35&-0.35&0.35\\
			$Q_{D}^{\prime}$ &0.9&0.1&0.5&-0.5&0.4\\
			$Q_{L}^{\prime}$ &0.15&0.05&0.75&-0.75&0.6\\
			$Q_{N}^{\prime}$ &-0.6&-0.4&-0.6&0.6&-0.4\\
			$Q_{E}^{\prime}$ &0.75&-0.3&-0.45&0.45&-0.35\\
			$Q_{H_u}^{\prime}$ &0.45&0.35&-0.15&0.15&-0.20\\
			$Q_{H_d}^{\prime}$ &-0.9&-0.05&-0.3&0.3&-0.25\\
			$Q_{S}^{\prime}$ &0.45&-0.3&0.45&-0.45&0.45\\
			$Q_{S_1}^{\prime}$ &0.45&-0.3&0.45&-0.45&0.45\\
			$Q_{S_2}^{\prime}$ &0.45&-0.3&0.45&-0.45&0.45\\
			$Q_{S_3}^{\prime}$ &-0.9&0.6&-0.9&0.9&-0.9\\
			\hline
\end{tabular}}
\label{tab:benchcharges}
\end{table}


\section{Conclusion}
\label{sec:conc}

{We realized that the $U(1)^{\prime}$ extension, which extends the MSSM with MSSM singlet particles considerably alter the DM phenomenology for the LSP neutralino masses from about 100 GeV to 1 TeV. This mass scales can be divided into three scenarios, which follow different manifestations in the results of the relic abundance of LSP neutralino and its scattering with nuclei. The typical mass spectrum in all scenarios involves two light CP-odd Higgs bosons whose masses are lighter than about 200 GeV and 600 GeV, respectively. Due to the strong impact from rare $B-$meson decays employed in our analyses, the MSSM Higgs bosons cannot be lighter than about 500 GeV, and such light Higgs boson solutions can be consistent with the constraints from rare $B-$meson decays only when they are formed mostly by the MSSM singlet fields. These CP-odd Higgs bosons play important roles to reduce the relic abundance of LSP neutralino compatible with the current Planck bounds.}

{Scenario I involves the LSP neutralino as light as about 100 GeV together with light charginos. It can be seen easily that the LSP composition involves more than about $80\%$ Singlino, while the remaining $ \lesssim 20\%$ also formed by the other MSSM singlet fields, $S_{1,2,3}$. This composition holds in almost all the solutions with $m_{\tilde{\chi}_{1}^{0}} \lesssim 500$ GeV. Thus, the secluded $U(1)^{\prime}$ model yields mostly MSSM singlet DM which can considerably interact with the MSSM particles through the Higgs portal. Singlino still takes part in the DM phenomenology for heavier LSP neutralino solutions, since its percentage in the LSP composition is realized greater than about $50\%$ for $m_{\tilde{\chi}_{1}^{0}} \lesssim 1$ TeV. In this region, the MSSM Higgsinos are involved in the LSP decomposition up to about $40\%$. The correct relic density for the LSP neutralino is partly satisfied through the chargino-neutralino coannihilation scenario in Scenario I up to about $m_{\tilde{\chi}_{1}^{\pm}}\lesssim 750$ GeV. The latest LHC constraints, especially the gluino mass bound, result in Binos and Winos heavier than about 300 GeV at the low scale SUSY spectrum, and hence the lighter chargino state is formed mostly by MSSM Higgsinos, so the chargino couples to the MSSM singlet LSP neutralino at tree-level. We realized that this coupling lead to a life-time for such light charginos always shorter than about $10^{-2}$ ns. We observed that the chargino life-time lasts longer if the Wino takes part in decomposition of the lighter chargino, but the solutions with $\tau_{\tilde{\chi}_{1}^{\pm}} \gtrsim 10^{-8}$ ns are excluded by the current Planck bound on the relic abundance of LSP neutralino. The solutions in Scenario II can also be characterized by the Singlino LSP neutralino in the similar mass range, but these solutions involve relatively heavier charginos, and the chargino-neutralino coannihilation scenario is not realized in this Scenario. The correct relic density of the LSP neutralino is satisfied through its annihilation into the lighter CP-odd Higgs bosons, which are mostly formed by the MSSM singlet Higgs bosons in our model.} 

{One may expect small scattering cross-section for the LSP neutralino due to its dominant singlet nature under the MSSM gauge group. However, as mentioned above, the Singlino is allowed to interact with the MSSM particles through the Higgs portal, which can potentially enhance its scattering cross-section. We observed that the Singlino LSP solutions can predict DM spin-independent cross-section in the range $\sim 4\times 10^{-46}- 10^{-50}$ cm$^2$ for the Singlino-like LSP. These solutions are expected to be tested in XENONnT experiments up to about $\sigma_{{\rm SI}}\sim 2\times 10^{-48}$ cm$^2$, while Darwin can lower the testable cross-section scale to about $3\times 10^{-49}$ cm$^2$ in near future.}

{We identify another class of solutions which can be classified as Scenario III. In this scenario, the LSP neutralino can still exhibit MSSM singlet nature, but the MSSM neutralinos considerably takes part in its decomposition since their mixing with the Singlino can be more than about $50\%$. Furthermore, this scenario can also yield MSSM-like LSP neutralino, which is mostly formed by the MSSM Higgsinos. The CMSSM-like gaugino mass relation yield $M_{2} > \mu$ and results in Higgsino-like lighter chargino states. Such solutions typically yield nearly mass degenerate chargino and LSP neutralino and they fall into the chargino-neutralino coannihilation region. The light CP-odd Higgs bosons do not take significant part in satisfying the correct relic density in this region. Even though such solutions can be consistent with the experimental constraints employed in our analyses, they receive a strong negative impact from the direct detection DM experiments due to the large scattering cross-sections of Higgsino-like LSP neutralino.  We find that its scattering cross-section is of the order about $10^{-44}$ cm$^{2}$, which is severely excluded by the current results from the direct detection experiments.}

\section{Acknowledgments} 
\label{sec:acknw}

The work of YH is supported by The Scientific and Technological Research Council of Turkey (TUBITAK) in the framework of the 2219-International Postdoctoral Research Fellowship Programme, and by Balikesir University Scientific Research Projects with grant No. BAP-2017/198. The research of C.S.U. was supported in part by the Spanish MICINN, under grant PID2019-107844GB-C22. The authors also acknowledge the use of the IRIDIS High Performance Computing Facility, and associated support services at the University of Southampton, in the completion of this work.

\bibliographystyle{JHEP}
\bibliography{Secluded}

\end{document}